\documentclass[prd, aps, twocolumn, showpacs,superscriptaddress,nofootinbib]{revtex4-1}
\usepackage[colorlinks=true,backref=true,linkcolor=blue,linktocpage=true,anchorcolor=black,citecolor=blue,filecolor=black,menucolor=black,pagecolor=black,urlcolor=blue]{hyperref}

\usepackage{graphicx,psfrag}
\usepackage{mathrsfs}
\usepackage{amsmath,amsfonts,amssymb}
\usepackage{multirow}
\usepackage{comment}
\usepackage{bm}
\usepackage{ulem}
\usepackage{hyperref}
\usepackage{relsize}
\usepackage{subfigure}

\newcommand{\be}{\begin{equation}}
\newcommand{\ee}{\end{equation}}
\newcommand{\bea}{\begin{eqnarray}}
\newcommand{\eea}{\end{eqnarray}}
\newcommand{\bel}{\begin{align}}
\newcommand{\eel}{\end{align}}

\def\p{\partial}

\def\GMc2{G M_{\odot} c^{-2}}

\def\O{\mathcal{O}}

\def\p{\partial}

\def\ii{{\rm i}}

\def\O{{\cal O}}

\def\TEOBResumS{\texttt{TEOBResumS}}

\def\SEOBNRvqT{{\texttt{SEOBNRv4T}}}

\DeclareSymbolFontAlphabet{\mathrsfs}{rsfs}
\DeclareMathAlphabet{\mathcal}{OMS}{cmsy}{m}{n}
\DeclareSymbolFontAlphabet{\mathrsfs}{rsfs}
\DeclareMathAlphabet\mathbfcal{OMS}{cmsy}{b}{n}

\usepackage{color}
\definecolor{cyan}{rgb}{0,0.9,0.9}
\definecolor{orange}{rgb}{0.9,0.5,0}
\definecolor{magenta}{rgb}{1,0,1}
\definecolor{purple}{rgb}{0.8,0.4,0.8}
\definecolor{gray}{rgb}{0.8242,0.8242,0.8242}
\definecolor{dodgerblue}{rgb}{0.12, 0.56, 1.0}

\begin{document}

\title{Nonlinear-in-spin effects in effective-one-body waveform models\\
  of spin-aligned, inspiralling, neutron star binaries}
\author{Alessandro \surname{Nagar}}
\affiliation{Centro Fermi - Museo Storico della Fisica e Centro Studi e Ricerche Enrico Fermi, Rome, Italy}
\affiliation{INFN Sezione di Torino, Via P. Giuria 1, 10125 Torino, Italy}
\affiliation{Institut des Hautes Etudes Scientifiques, 91440 Bures-sur-Yvette, France}
\author{Francesco \surname{Messina}}
\affiliation{Dipartimento di Fisica, Universit\`a degli studi di Milano Bicocca, Piazza della Scienza 3, 20126 Milano, Italy}
\affiliation{INFN, Sezione di Milano Bicocca, Piazza della Scienza 3, 20126 Milano, Italy}
\author{Piero \surname{Rettegno}}
\affiliation{INFN Sezione di Torino, Via P. Giuria 1, 10125 Torino, Italy}
\affiliation{Dipartimento di Fisica, Universit\`a di Torino, via P. Giuria 1, 10125 Torino, Italy}
\author{Donato \surname{Bini}}
\affiliation{Istituto per le Applicazioni del Calcolo \lq\lq M.~Picone", CNR, I-00185 Rome, Italy}
\affiliation{INFN, Sezione di Roma Tre, I-00146 Roma, Italy}
\author{Thibault \surname{Damour}}
\affiliation{Institut des Hautes Etudes Scientifiques, 91440 Bures-sur-Yvette, France}
\author{Andrea \surname{Geralico}}
\affiliation{Istituto per le Applicazioni del Calcolo \lq\lq M.~Picone", CNR, I-00185 Rome, Italy}
\author{Sarp \surname{Akcay}}
\affiliation{Theoretisch-Physikalisches Institut, Friedrich-Schiller-Universit{\"a}t Jena, 07743, Jena, Germany}  
\author{Sebastiano \surname{Bernuzzi}}
\affiliation{Theoretisch-Physikalisches Institut, Friedrich-Schiller-Universit{\"a}t Jena, 07743, Jena, Germany}  
\begin{abstract}
Spinning neutron stars acquire a quadrupole moment due to their own rotation. This quadratic-in-spin, self-spin effect
depends on the equation of state (EOS) and affects the orbital motion and rate of inspiral of
neutron star binaries. Building upon circularized post-Newtonian results, we incorporate the EOS-dependent
self-spin (or monopole-quadrupole)  terms in the spin-aligned effective-one-body (EOB) waveform model \TEOBResumS{} 
at next-to-next-to-leading (NNLO) order, together with other (bilinear, cubic and quartic) nonlinear-in-spin effects (at leading order, LO). 
We point out that the structure of the Hamiltonian of \TEOBResumS{} is such  that it already incorporates,
in the binary black hole case, the recently computed [Levi and Steinhoff, arXiv:1607.04252] quartic-in-spin LO term.
Using the gauge-invariant characterization of the phasing provided by the function 
$Q_\omega=\omega^2/\dot{\omega}$ of $\omega=2\pi f$ , where $f$ is the gravitational wave frequency, 
we study the EOS dependence of the self-spin effects and show that: 
(i) the next-to-leading order (NLO) and NNLO monopole-quadrupole corrections yield
increasingly phase-accelerating effects compared to the corresponding LO contribution;
(ii) the standard TaylorF2 post-Newtonian (PN) treatment of NLO (3PN)  EOS-dependent self-spin 
effects makes their action stronger than the corresponding EOB description; 
(iii) the addition to the standard 3PN TaylorF2 post-Newtonian phasing description 
of self-spin tail effects at LO allows one to reconcile the self-spin part of the TaylorF2 PN 
phasing with the corresponding \TEOBResumS{} one up to dimensionless frequencies  
$M\omega\simeq 0.04-0.06$. Such a tail-augmented TaylorF2 approximant then yields  
an analytically simplified, EOB-faithful, representation of the EOS-dependent self-spin 
phasing that can be useful to improve current PN-based (or phenomenological) waveform 
models  for inspiralling neutron star binaries.
Finally, by generating the inspiral dynamics using the post-adiabatic approximation, incorporated
in a new implementation of \TEOBResumS{}, one finds that the computational time needed to obtain
a typical waveform (including all multipoles up to $\ell=8$) from 10~Hz is of the order of 0.4~sec.
\end{abstract}

\maketitle

\section{Introduction}
Neutron stars (NSs) are self-gravitating bodies inside which matter is compressed to very high densities.
Gravitational wave (GW) signals can be used to put constraints on the equation of state
(EOS) of degenerate matter in these extreme environments.
In fact, when a NS is part of a binary system, their mutual tidal 
interaction deform the stars, affecting the dynamics of the system and the emitted GWs.
On August 17, 2017, the first binary neutron star (BNS) inspiral has been detected by
the LIGO-Virgo interferometers~\cite{TheLIGOScientific:2017qsa}.
One of the important outcomes of this discovery was the measurement of the neutron
star radii and EOS from the GW signal~\cite{Abbott:2018wiz,Abbott:2018exr}
obtained by extracting from the data the tidal polarizabilities (or deformabilities)
related to the NS Love numbers~\cite{Damour:1983a,Flanagan:2007ix,Damour:2009vw,Damour:2012yf}.

When NSs are spinning, the rate of the inspiral can be modified by an additional 
EOS-dependent effect, since each NS acquires a quadrupole moment due to its own rotation. 
The importance of such spin-induced-monopole-quadrupole effects on BNS inspirals 
was pointed out long ago~\cite{Poisson:1997ha} and recently
revived~\cite{Harry:2018hke} in a data-analysis context, emphasizing that it is important to
incorporate such self-spin terms in BNS waveform templates to avoid parameter biases in the
case of highly spinning BNS systems. In addition, it was also recently pointed out
that self-spin effects might be useful to test the binary black hole nature
of the compact objects~\cite{Krishnendu:2017shb,Krishnendu:2018nqa}. 
Consistently with these findings, the analysis of GW170817 
was done with waveform models that {\it do} include
EOS-dependent self-spin effects. These were incorporated in resummed form in the
\SEOBNRvqT{}~\cite{Steinhoff:2016rfi} and \TEOBResumS{}~\cite{Nagar:2018zoe}
effective one body (EOB) models and in TaylorF2-like post-Newtonian (PN)
form in the {\tt PhenomPv2NRTidal} model~\cite{Dietrich:2017aum,Dietrich:2018uni}.
Both descriptions have their drawbacks and can be improved.
On the one hand, the {\tt PhenomPv2NRTidal} description is incorporating self-spin terms up to next-to-leading
order (NLO), but it is biased by the fact that the PN approximation breaks down at some stage 
in the relativistic regime close to merger. On the other hand, the EOB description is robust up to merger, 
but only the leading order (LO) self-spin effects (both in the EOB Hamiltonian and flux) were
included in the models. Although one of the main results of Ref.~\cite{Dietrich:2018uni} was to
show good consistency between \TEOBResumS{} and {\tt PhenomPv2NRTidal}, this was not a precise
apple-with-apple comparison because of the additional NLO self-spin effects included in {\tt PhenomPv2NRTidal}
and not in \TEOBResumS{}. Furthermore \TEOBResumS{} is actually taking into account an infinite
number of self-spin tail terms (in the waveform and flux), that are absent
in {\tt PhenomPv2NRTidal}, so that the precise question about which model is 
analytically more complete requires an elaborate study.
In particular, none of the current waveform models that use a 3.5PN-accurate inspiral description
(like TaylorF2 or {\tt PhenomPv2NRTidal}~\cite{Dietrich:2018uni}) are using the EOS-dependent
3.5PN-accurate self-spin tail term, although it is available analytically~\cite{Nagar:2018zoe}.
Such a term can be obtained by suitably expanding the EOB energy and flux along circular orbits,
adapting the procedure of Ref.~\cite{Messina:2017yjg} (see also~\cite{Damour:2012yf}), 
that allowed one to cross check the 4.5PN, nonspinning, tail term in the flux formerly obtained 
from an \textit{ab initio} PN calculation~\cite{Marchand:2016vox}.
Finally, we mention that state-of-the-art NR simulations of coalescing
BNSs~\cite{Bernuzzi:2011aq,Bernuzzi:2012ci,Radice:2013hxh,Bernuzzi:2014owa,Radice:2015nva,Bernuzzi:2016pie,Radice:2016gym}
are currently barely able to resolve spin-quadratic effects close to merger~\cite{Bernuzzi:2013rza,Dietrich:2016lyp}
and are too short to measure their cumulative effect during many inspiral orbits.
As a consequence, we can only rely on analytical models for their description
for LIGO/Virgo targeted analyses.

The purpose of this paper is then to address and answer the questions that remained open in
Refs.~\cite{Nagar:2018zoe,Dietrich:2018uni}. We do so by extending the EOS-dependent
self-spin sector of \TEOBResumS{} to NLO and next-to-next-to-leading order (NNLO),
suitably recasting in EOB form recent PN calculations of Levi and
Steinhoff~\cite{Levi:2015ixa,Levi:2015uxa,Levi:2016ofk}. For simplicity, this is done
in the circular approximation by exploiting the gauge-invariant relation between
energy and angular momentum rather than by deriving the explicit canonical transformation
that maps the Arnowitt-Deser-Misner (ADM) Hamiltonian~\cite{Schafer:2018kuf} 
into the EOB Hamiltonian. This new knowledge allows us to produce a 
consistent phasing comparison with the TaylorF2 approximant.
We find that the phase accelerating effect of the spin-induced quadrupole moment terms
is {\it enhanced} by the NLO contribution, although the magnitude of the effect
as predicted by \TEOBResumS{} is always {\it smaller} than in the corresponding
TaylorF2 description. Remarkably, a TaylorF2 approximant
that also incorporates the LO self-spin tail effect yields a self-spin phasing
that is essentially equivalent to the NLO \TEOBResumS{} one up to
frequency $M\omega\simeq 0.05$ independently of the EOS choice.
We also show that the LO quartic-in-spin effects 
entering the circularized Hamiltonian recently computed by Levi and Steinhoff~\cite{Levi:2016ofk}
are already contained in the \TEOBResumS{} Hamiltonian of Ref.~\cite{Damour:2014sva}
 in the black-hole (BH) case, due to the use of the centrifugal radius.
The corresponding correction to the centrifugal radius yielded by the
octupolar and hexadecapolar EOS-dependent spin-induced effects (in the non binary BH case)
is explicitly obtained. 

The paper is organized as follows: Section~\ref{sec:summary} builds upon Ref.~\cite{Nagar:2018zoe}
and describes how spin-quadratic (and spin-quartic) terms are incorporated in the
Hamiltonian of \TEOBResumS, computing the additional corrections to the centrifugal
radius $r_c$~\cite{Damour:2014sva}. Section~\ref{sec:PN} summarizes the spin sector
of the, closed form, frequency domain, waveform approximant TaylorF2. The predictions
of \TEOBResumS{} and of TaylorF2 for what concerns the monopole-quadrupole effects are
compared in Sec.~\ref{sec:results}. Section~\ref{sec:conclusions} collects some concluding
remarks. The paper ends with three  appendices: Appendix~\ref{app:S3} presents a few suggestions
about how incorporating the cubic-in-spin dynamical effects of~\cite{Levi:2016ofk} within the EOB Hamiltonian;
Appendix~\ref{sec:PA} illustrates the performance of the (high-order) post-adiabatic dynamics,
as discussed in Ref.~\cite{Nagar:2018gnk}, to efficiently compute long-inspiral
BNS waveform. In particular, we find that the computational time needed for generating a
typical (time-domain) BNS waveform (summed over all multipoles up to $\ell=8$ inclueded)
from 10~Hz is of the order of 0.4, that goes down to 0.1~sec from 20~Hz.
Appendix~\ref{app:S_conv} re-expresses our results in terms of different spin variables.
If not otherwise specified, we use units with $G=c=1$.

%===============================================
\section{Nonlinear-in-spin effects within TEOBResumS}
\label{sec:summary}
%===============================================
The EOS-dependent self-spin contribution at LO in \TEOBResumS{} was discussed extensively
in Sec.~IIIB of Ref.~\cite{Nagar:2018zoe} to which we refer the reader for further details.
Our notation follows~\cite{Nagar:2018zoe}. We consider binary systems in which the
two bodies are labeled by $(A,B)$. Their masses and dimensional spins are 
denoted  $M_{A,B}$ (with $M_A \geq M_B$) and $S_{A,B} \equiv M_{A,B} a_{A,B}$ respectively.
The total mass is $M = M_A + M_B$ and the reduced mass $\mu = (M_A M_B)/M$.
We also introduce the mass ratio $q \equiv M_A/M_B \geq 1$, the symmetric mass ratio $\nu = \mu/M$,
the mass fractions $X_{A,B}\equiv M_{A,B}/M$ and the shorthand $X_{AB}\equiv X_A-X_B = \sqrt{1-4\nu}$.
Finally, we make use of the dimensionless spin variables $\tilde{a}_i \equiv a_i/M \equiv S_i/(M_i M)$ 
together with their symmetric and antisymmetric 
combinations\footnote{Note the difference between $\tilde{a}_i \equiv a_i/M$ and the usually introduced
dimensionless spin $\hat{a}_i \equiv \chi_i \equiv a_i/M_i$.  Note also that in Refs~\cite{Nagar:2016ayt,Messina:2018ghh,Nagar:2018zoe}
we had denoted $\tilde{a}_0$ as $\hat{a}_0$.}  $\tilde{a}_0 = \tilde{a}_A + \tilde{a}_B$
and $\tilde{a}_{AB} = \tilde{a}_A - \tilde{a}_B$.

\subsection{Hamiltonian: quadratic-in-spin terms}
\label{sec:S2}
In \TEOBResumS{}~\cite{Damour:2014yha,Nagar:2018zoe}, which is limited to
the case of spin-aligned (nonprecessing) binaries, spin-quadratic effects
are treated introducing the ``centrifugal radius" $r_c$, considered as a function of the Boyer-Lindquist-type
EOB radial variable $r$, and of the spin variables.
For BBHs, the function $r_c(r, \tilde{a}_A, \tilde{a}_B)$ incorporates both LO and NLO spin-quadratic effects~\cite{Hartung:2010jg,Balmelli:2015lva,Damour:2014yha};
by contrast only LO spin-quadratic effects were considered for
extended objects like NSs~\cite{Nagar:2018zoe}.
We hence start by generalizing the expression of the centrifugal radius in
order to take into account both NLO and NNLO, EOS-dependent, self-spin
effects, exploiting the PN-expanded results of Refs.~\cite{Levi:2015ixa,Levi:2015uxa,Levi:2016ofk}.
The generalized formula for the centrifugal radius that formally takes into
account both NLO and NNLO spin-quadratic effects reads
\begin{equation} \label{rc2}
r_c^2(r,\tilde{a}_A,\tilde{a}_B)^{\rm NNLO}=r^2+\tilde{a}_{Q}^2\left(1+\frac{2}{r}\right) +\frac{\delta a^2_{\rm NLO}}{r} +\frac{\delta a_{\rm NNLO}^2}{r^2},
\end{equation}
where we are using a dimensionless radial coordinate $r \equiv \frac{R}{M}$, and we
introduced the effective spin variable
\begin{equation} \label{aQ}
\tilde{a}_{Q}^2 \equiv C_{QA}\tilde{a}_A^2+2\tilde{a}_A\tilde{a}_B+C_{QB}\tilde{a}_B^2.
\end{equation}
$C_{QA}$ and $C_{QB}$ are coefficients that parametrize the quadrupolar deformation
acquired by the NSs due to their own rotation. For binary black holes,
$C_{Qi} = 1$, so $\tilde{a}_{Q}^2$ reduces to $\tilde{a}_0^2$.
The parameters $\delta a^2_{\rm NLO}$ and $\delta a_{\rm NNLO}^2$ encode the NLO
and NNLO spin-spin information respectively.
As mentioned above, working in the circular approximation for simplicity,
we compute them exploiting the functional relation between binding energy and
orbital angular momentum, that is explicitly given, in PN-expanded form, in
Refs.~\cite{Levi:2015ixa,Levi:2015uxa,Levi:2016ofk}.
In practice, one computes  the PN-expanded EOB dynamics along circular orbits,
that will explicitly depend on $(\delta a^2_{\rm NLO},\delta a_{\rm NNLO}^2)$, 
and then fixes these coefficients by comparison with the PN-expanded relation
of Ref.~\cite{Levi:2016ofk}.

To do so, let us recall the main elements of the Hamiltonian of \TEOBResumS{}
that are useful for this calculation. Since we are considering 
nonprecessing systems, the dynamics is described by the dimensionless
phase-space variables $(r,p_{r_*},\varphi,p_\varphi)$. We use $\varphi$ to denote the orbital phase,
while the (dimensionless) radial and angular momentum are respectively 
defined as $p_{r_*}=P_{R_*}/\mu$ and $p_\varphi = P_\varphi/(\mu M)$. 
The $\mu$-rescaled EOB Hamiltonian then reads
\begin{equation}
\hat{H}_{\rm EOB} = \frac{H_{\rm EOB}}{\mu} = \frac1\nu \sqrt{1+2\nu\big(\hat{H}_{\rm eff}-1\big)},
\end{equation}
where $\hat{H}_{\rm eff} = \hat{H}_{\rm eff}^{\rm orb} + p_\varphi \tilde{G}$, i.e.,
the sum of a orbital (even in spin) and spin-orbit (odd in spin) term.
Here $\tilde{G}$ explicitly reads
\be
\label{eq:G}
\tilde{G}= G_S \hat{S} + G_{S_*} \hat{S}_*,
\ee
where $\hat{S}\equiv (S_A + S_B)/M^2$ and $\hat{S}_* \equiv [(M_B/M_A) S_A+(M_A/M_B) S_B]/M^2$.
As in previous work, the functions $(G_S,G_{S_*})$ are written in
Damour-Jaranowski-Sch\"afer
gauge~\cite{Damour:2008qf,Nagar:2011fx}, which means gauging away the dependence
on the angular momentum $p_\varphi$ so that they  depend only on $u \equiv 1/r$
and on $p_{r_*}$. The explicit expressions of $(G_S,G_{S_*})$ can be found
in Refs.~\cite{Nagar:2011fx,Damour:2014yha}. These expression only retain, in the
spin-orbit part of the Hamiltonian, terms that are {\it linear} in the spins.
However, the complete \TEOBResumS{} model is based on the prescription of
Refs.~\cite{Damour:2014sva,Nagar:2015xqa,Nagar:2018zoe} to effectively incorporate,
in resummed form, also higher odd-powers of the spins (spin-cubed, spin$^5$ etc.)
by suitably replacing the $u$-dependence of the functions $(G_S,G_{S_*})$ with
dependence on $u_c$. We shall see in Sec.~\ref{sec:Ham_s3} below
that \TEOBResumS{} delivers a reasonable approximation to the actual LO
spin-cubic part of the ADM Hamiltonian of Ref.~\cite{Levi:2016ofk}.
In Appendix~\ref{app:S3} we give possible EOB transcriptions
of the results of~\cite{Levi:2016ofk}.  

The orbital part of the effective Hamiltonian reads
\begin{equation}
\hat{H}_{\rm eff}^{\rm orb} = \sqrt{p_{r_*}^2+A \left(1+p_\varphi^2 u_c^2+z_3 ~p_{r_*}^4 u_c^2\right)},
\end{equation}
with $z_3 = 2\nu(4-3\nu)$ and $u_c \equiv 1/r_c$.
$A$ is the effective metric potential, whose PN expansion in the non-spinning limit is
\begin{align}
\label{PN_metric}
A^{\rm PN}_{\rm orb} (u) =&~ 1 -2 u + 2 \nu u^3 + \nu a_4 u^4 \nonumber \\
&+ \nu \big(a_5^c + a_5^{\log}\log(u)\big)u^5 + \O[u^6].
\end{align}
The PN coefficients that appear above explicitly read
\begin{align}
a_4~ =&~ \frac{94}{3} - \frac{41}{32} \pi^2, \nonumber \\
a_5^c~ =& -\frac{4237}{60} + \frac{2275}{512} \pi^2 + \frac{256}{5} \log 2 + \frac{128}{5} \gamma \nonumber \\
&- \left(\frac{221}{6} - \frac{41}{32} \pi^2\right)\nu , \nonumber \\
a_5^{\log} =&~ \frac{64}{5},
\end{align}
where $\gamma=0.57721\dots$ is Euler's constant.
In \TEOBResumS{}, this effective metric is resummed using a Pad\'e approximant, namely 
\begin{equation}
A_{\rm orb} (u) = {\rm P}^1_5 [ A^{\rm PN}_{\rm orb}(u)].
\end{equation}
When spins are present, the metric is built upon the Kerr one and reads
\begin{equation}
A (u; S_i) =~ \frac{1 + 2u_c}{1 + 2u} A_{\rm orb} (u_c).
\end{equation}
We hence start considering circular orbits ($p_{r_*}=0$) and compute the circular angular momentum, $j$, using the condition $\partial_u\hat{H}_{\rm eff}=0$, that yields
the following equation
\begin{align}
\label{j0_circ}
&\left\{\left[\left(A u_c^2\right)'\right]^2 - 4 A u_c^2 \left(\tilde{G}'\right)^2\right\}j^4 \nonumber \\
&+\left[2 A' \left(A u_c^2\right)' - 4 A \left(\tilde{G}'\right)^2\right]j^2 + \left(A'\right)^2 = 0,
\end{align}
where the prime indicates $(\cdot)^\prime \equiv \p_u(\cdot)$. 
By expanding the solution of Eq.~\eqref{j0_circ} in series of $u$ and up to the second order in spin one obtains
\begin{widetext}
\begin{align}
\label{eq:j0}
j(u)=&~
\frac{1}{\sqrt{u}}
+\frac{3}{2}\sqrt{u}
-\frac{3}{8}\left( 7 \tilde{a}_0 + X_{AB} \tilde{a}_{AB}\right) u+\biggl[\frac{27}{8}-\frac{3}{2}\nu+\tilde{a}_{Q}^2\biggr]u^{3/2}
\nonumber\\
&+\left[\left(-\frac{87}{16}+\frac{11}{8}\nu\right)\tilde{a}_0 - \left(\frac{33}{16}+\frac{1}{8}\nu\right)X_{AB}\tilde{a}_{AB}\right]u^2\nonumber\\
&+\biggl[\frac{135}{16}+\left(-\frac{433}{12}+\frac{41}{32}\pi^2\right)\nu+ \frac{441}{128}\tilde{a}_0^2 +
\frac12 \tilde{a}_{Q}^2 + \left(\frac{9}{128}-\frac{9}{32}\nu\right)\tilde{a}_{AB}^2 +\frac{63}{64}X_{AB}\tilde{a}_0\tilde{a}_{AB}
+\frac{5}{4}\delta a_{\rm NLO}^2\biggr]u^{5/2} \nonumber\\
& +\left[\left(-\frac{63}{4}+\frac{505}{16}\nu +\frac{25}{64}\nu^2\right)\tilde{a}_0 + \left(-\frac{63}{8}+\frac{55}{16}\nu -\frac{5}{64}\nu^2\right)X_{AB}\tilde{a}_{AB}\right]u^3 \nonumber \\
&+\Biggl[\frac{2835}{128}-\left(\frac{3029}{120} + 32 \gamma + \frac{3503}{2048}\pi^2 + 64 {\rm log}(2) + 16 {\rm log}(u)  \right)\nu + \left(\frac{539}{12}-\frac{205}{128}\pi^2\right)\nu^2 \nonumber \\
&+ \left(\frac{4095}{256}-\frac{231}{64}\nu\right)\tilde{a}_0^2-\left(\frac{9}{8}-\frac{9}{4}\nu\right)\tilde{a}_{Q}^2 +\left(\frac{207}{256}-\frac{51}{16}\nu-\frac{3}{16}\nu^2\right)\tilde{a}_{AB}^2 \nonumber \\
&+ \left(\frac{1017}{128}-\frac{3}{16}\nu\right)X_{AB}\tilde{a}_0\tilde{a}_{AB} - \frac{19}{8}\delta a_{\rm NLO}^2
 + \frac{3}{2}\delta a_{\rm NNLO}^2 \Biggr]u^{7/2} + \O\left[u^4\right].
\end{align}
\end{widetext}
 This truncated series can be inverted so to obtain $u(j)$,
which reads
\begin{widetext}
\begin{align}
\label{eq:uj}
u(j)=&~\frac{1}{j^2}+\frac{3}{j^4}-\frac{3}{4}\left(7\tilde{a}_0+X_{AB}\tilde{a}_{AB}\right)\frac{1}{j^5}
+\left(18-3\nu+2\tilde{a}_{Q}^2\right)\frac{1}{j^6}\nonumber\\
&+\left[\left(-\frac{465}{8} +\frac{11}{4}\nu\right)\tilde{a}_0 - \left(\frac{87}{8}+\frac{\nu}{4}\right)X_{AB}\tilde{a}_{AB}\right]\frac{1}{j^7}\nonumber\\
&+\biggl[135+\left(-\frac{311}{3}+\frac{41}{16}\pi^2\right)\nu +\frac{441}{8}\tilde{a}_0^2 + 22\tilde{a}_{Q}^2 + \left(\frac{9}{8}-\frac{9}{2}\nu \right)\tilde{a}_{AB}^2+\frac{63}{4}X_{AB}\tilde{a}_0\tilde{a}_{AB}+\frac{5}{2}\delta a_{\rm NLO}^2\biggr]\frac{1}{j^8} \nonumber \\
&+ \left[\left(-\frac{1269}{2}+\frac{1273}{8}\nu+\frac{25}{32}\nu^2\right)\tilde{a}_0+\left(-\frac{531}{4}+\frac{103}{8}\nu-\frac{5}{32}\nu^2\right)X_{AB}\tilde{a}_{AB}\right] \frac{1}{j^9} \nonumber \\
&+\biggl[1134-\left(\frac{163063}{120}+64 \gamma- \frac{31921}{1024}\pi^2 + 128 {\rm log}(2) + 64 {\rm log}\left(1/j\right) \right)\nu + \left(\frac{1321}{12}- \frac{205}{64}\pi^2 \right)\nu^2 \nonumber \\
&+ \left(\frac{9009}{8}-\frac{1155}{16}\nu\right)\tilde{a}_0^2
+ \left(234-\frac{45}{2}\nu\right)\tilde{a}_{Q}^2 + \left(\frac{261}{8}-\frac{2073}{16}\nu -\frac{15}{4}\nu^2 \right)\tilde{a}_{AB}^2 \nonumber \\
&+ \left(\frac{1557}{4}-\frac{15}{4}\nu\right)X_{AB}\tilde{a}_0\tilde{a}_{AB} + 29~\delta a_{\rm NLO}^2 + 3~\delta a_{\rm NNLO}^2\biggr]\frac{1}{j^{10}}+ \O\left[j^{-11}\right].
\end{align}
\end{widetext}
By placing this expanded expression of $u$ into the EOB Hamiltonian,
one can finally obtain the gauge-invariant relation between the binding
energy and angular momentum. The binding energy per reduced mass is in fact
defined as $E_b=(E-M)/\mu$, where $E=\nu \hat{H}_{\rm EOB}$,
and is given as a polynomial in inverse powers of $j$, i.e.,
\begin{equation}
E_b(j) = -\dfrac{1}{2j^2}\left(1+\sum_{n=1}^{8}\frac{c_n}{j^n} + \O\left[j^{-9}\right]\right).
\end{equation}
Explicitly, from the expansion of the EOB Hamiltonian along circular orbits
we get
\begin{widetext}
\begin{align}
\label{eq:Ej}
E_b(j)=&-\frac{1}{2 j^2} \Bigg\{1
+\frac{1}{4}(9+\nu)\frac{1}{j^2}
-\frac{1}{2}\left(7\tilde{a}_0+X_{AB}\tilde{a}_{AB}\right)\frac{1}{j^3}\nonumber\\
&
+\frac{1}{8}\biggl[81-7\nu+\nu^2+8\tilde{a}_{Q}^2\biggr]\frac{1}{j^4}
-\frac{3}{8}\left[\left(81+\nu\right)\tilde{a}_0 + \left(15+\nu\right)X_{AB}\tilde{a}_{AB}\right]\frac{1}{j^5}\nonumber\\
&+\biggl[\frac{3861}{64}-\left(\frac{8833}{192}-\frac{41}{32}\pi^2\right)\nu-\frac{5}{32}\nu^2+\frac{5}{64}\nu^3+\frac{441}{16}\tilde{a}_{0}^2+\left(\frac{17}{2}+\frac{\nu}{2}\right) \tilde{a}_{Q}^2+\left(\frac{9}{16}-\frac{9}{4}\nu \right)\tilde{a}_{AB}^2\nonumber \\
&+\frac{63}{8}X_{AB}\tilde{a}_0\tilde{a}_{AB}+\delta a^2_{\rm NLO}\biggr]\frac{1}{j^6}
+\frac{1}{16}\left[\left(-4293+822\nu-5\nu^2\right)\tilde{a}_0+\left(-891+42\nu-5\nu^2\right)X_{AB}\tilde{a}_{AB}\right]\frac{1}{j^7}\nonumber \\
&+\biggl[\frac{53703}{128}-\left(\frac{989911}{1920}+\frac{128}{5} \gamma -\frac{6581}{512}\pi^2 +\frac{256}{5} {\rm log}(2) + \frac{128}{5} {\rm log}\left(1/j\right) \right)\nu + \left(\frac{8875}{384}- \frac{41}{64}\pi^2 \right)\nu^2 \nonumber \\
&-\frac{3}{64}\nu^3 +\frac{7}{128}\nu^4+ \left(\frac{14679}{32}-\frac{385}{32}\nu\right)\tilde{a}_0^2
+\left(\frac{603}{8}-\frac{29}{8}\nu+\frac{3}{8}\nu^2\right)\tilde{a}_Q^2
+\left(\frac{423}{32}-\frac{1669}{32}\nu -\frac{23}{8}\nu^2 \right)\tilde{a}_{AB}^2 \nonumber \\
&+\left(\frac{2529}{16}+\frac{53}{16}\nu\right)X_{AB}\tilde{a}_0\tilde{a}_{AB}+\left(\frac{19}{2}+\frac{\nu}{2}\right)\delta a_{\rm NLO}^2 + \delta a_{\rm NNLO}^2\biggr]\frac{1}{j^{8}}+ \O\left[j^{-9}\right]\Bigg\},
\end{align}
\end{widetext}
where we see that the $c_6$ and $c_8$ coefficients explicitly depend on $\delta a_{\rm NLO}^2$ and $\delta a_{\rm NNLO}^2$.
The corresponding quantities in Eq.~(5.3) of Ref.~\cite{Levi:2015uxa}, once expressed in our spin variables\footnote{Note that 
Ref.~\cite{Levi:2015uxa} uses as dimensionless spin variables some quantities, $S_i^{\rm L-S}$,
that correspond to our $S_i/(M_A M_B)$. Furthermore, their
deformation coefficients are denoted by $(C_{ES^2},C_{BS^3},C_{ES^4})$, in order to highlight the spin order
and their electric/magnetic behavior. In our convention, they correspond to $(C_Q,C_\text{Oct},C_\text{Hex})$
respectively, which puts the accent on the multipole of the deformation. We also note that the $\lambda$ constants by Marsat (see Sec. B of Ref. \cite{Marsat:2014xea}) are the same as Levi and Steinhoff's $C_{BS^3}$ and our $C_\text{Oct}$'s.}, explicitly read
\begin{widetext}
\begin{align}
\label{j8levi}
\left(c_6^{\rm SS}\right)^{\rm L-S}=&~ \frac{1}{16}\Biggl\{\left(375+8\nu\right)\tilde{a}_0^2 +8\left(-23+\nu\right)\tilde{a}_{Q}^2 +\left(7-52\nu\right)\tilde{a}_{AB}^2 + X_{AB}\left[ 130~\tilde{a}_0 \tilde{a}_{AB} + 16\left(C_{QA}\tilde{a}_A^2-C_{QB}\tilde{a}_B^2\right)\right]  \Biggr\}, \\
\label{j10levi}
\left(c_8^{\rm SS}\right)^{\rm L-S} =&~ \frac{1}{112}\Biggl\{-\left(51369-2743\nu+21\nu^2\right)\tilde{a}_0^2 +\left(13182-1066\nu+42\nu^2\right)\tilde{a}_{Q}^2 -\left(5205+6292\nu+329\nu^2\right)\tilde{a}_{AB}^2 \nonumber \\
&+ X_{AB}\left[13\left(1380+7\nu\right)\tilde{a}_0 \tilde{a}_{AB} + \left(1716+56\nu\right)\left(C_{QA}\tilde{a}_A^2-C_{QB}\tilde{a}_B^2\right)\right]  \Biggr\}.
\end{align}
\end{widetext}

Comparing Eqs.~\eqref{j8levi} and \eqref{j10levi} to Eq.~\eqref{eq:Ej} one obtains
\begin{align}
\label{deltaa2}
\delta a^2_{\rm NLO} =&-\frac{33}{8}\tilde{a}_0^2 + 3\tilde{a}_{Q}^2 -\frac{1}{8}\left(1+4\nu\right)\tilde{a}_{AB}^2\nonumber \\
&+X_{AB} \left[\frac{1}{4}\tilde{a}_0\tilde{a}_{AB} + \left(C_{QA}\tilde{a}_A^2-C_{QB}\tilde{a}_B^2\right)\right],
\end{align}
and
\begin{align}
\delta a_{\rm  NNLO}^2 =& -\left(\frac{4419}{224}+\frac{1263}{224}\nu\right)\tilde{a}_0^2 + \left(\frac{387}{28}-\frac{207}{28}\nu \right) \tilde{a}_{Q}^2 \nonumber \\
&+\left(\frac{11}{32}-\frac{127}{32}\nu+\frac{3}{8}\nu^2\right)\tilde{a}_{AB}^2 \nonumber \\
&+X_{AB} \biggl[-\left(\frac{29}{112}+\frac{21}{8}\nu\right)\tilde{a}_0\tilde{a}_{AB} \nonumber \\
&+\frac{163}{28}\left(C_{QA}\tilde{a}_A^2-C_{QB}\tilde{a}_B^2\right)\biggr].
\end{align}

\subsubsection{Binary black hole limit}
The BH case is recovered imposing $C_{QA}=C_{QB}=1$ or, equivalently, $\tilde{a}^2_{Q} = \tilde{a}_0^2$ and $\left(C_{QA}\tilde{a}_A^2-C_{QB}\tilde{a}_B^2\right) = \tilde{a}_0 \tilde{a}_{AB}$. This yields
\begin{align}
\label{deltaa2BH}
\delta a^2_{\rm BBH \,  NLO} =&-\frac{9}{8}\tilde{a}_0^2 -\frac{1}{8}\left(1+4\nu\right)\tilde{a}_{AB}^2 + \frac{5}{4}X_{AB} \tilde{a}_0\tilde{a}_{AB} ,\\
\delta a_{\rm BBH \,  NNLO}^2 =&-\left(\frac{189}{32}+\frac{417}{32}\nu\right)\tilde{a}_0^2 \nonumber \\
&+\left(\frac{11}{32} - \frac{127}{32}\nu + \frac{3}{8}\nu^2\right)\tilde{a}_{AB}^2 \nonumber \\
&+\left(\frac{89}{16}-\frac{21}{8}\nu\right)X_{AB}\tilde{a}_0\tilde{a}_{AB}. 
\end{align}
Eq.~\eqref{deltaa2BH} agrees with the result for the same quantity obtained in
Ref.~\cite{Damour:2014yha} (see Eq.~(60) there) with a different method
(see also~\cite{Balmelli:2015zsa}). The impact of the newly computed
$\delta a_{\rm NNLO}^2$ in BBH systems will be analyzed elsewhere.

\subsection{Hamiltonian: cubic-in-spin terms already included in \TEOBResumS{}}
\label{sec:Ham_s3}
The LO cubic-in-spin contribution to the PN-expanded Hamiltonian
(and thus on the $E_b(j)$ curve) was derived in Ref.~\cite{Levi:2015uxa}.
This contribution is not fully incorporated in the current version
of \TEOBResumS{}. However, one should be aware that some cubic-in-spin
terms {\it are already included} in the model, because they naturally
arise due to the presence of $u_c$ in the gyro-gravitomagnetic
functions $G_S$ and $G_{S_*}$ that enter the spin-orbit sector of the
Hamiltonian (see Sec.~\ref{sec:S2}). It is then interesting to check how
these terms, that are guessed by the resummed structure of the Hamiltonian,
do compare with the exact result of Ref.~\cite{Levi:2015uxa}.
We now redo the calculations of Sec.~\ref{sec:S2}, this time keeping the cubic-in-spin terms, whose LO enters in the coefficient of $j^{-9}$ in the gauge-invariant relation $E_b (j)$.
In \TEOBResumS{}, the former is given by
\begin{equation}
\label{EjS3}
\left(c_{7}^{S^3_{\rm LO}}\right)^{\TEOBResumS{}} = 
-\left(\frac{67}{4}\tilde{a}_0 + \frac{13}{4} X_{AB} \tilde{a}_{AB}\right)\tilde{a}_Q^2.
\end{equation}
By contrast, the PN-expanded result from Ref.~\cite{Levi:2015uxa} reads
\begin{align}
\label{EjS3_LS}
\left(c_{7}^{S^3_{\rm LO}}\right)^{\rm L-S} =& -\left(2 C_{\text{Oct} A} + 15 C_{Q A} + 3 X_{AB} C_{Q A}\right)\tilde{a}_A^3 \nonumber \\
&- \left[30 + 21 C_{Q A} + X_{AB}\left(6 -3C_{Q A}\right) \right]\tilde{a}_A^2\tilde{a}_B \nonumber \\
&- \left[30 + 21 C_{Q B} - X_{AB}\left(6-3C_{Q B}\right)\right]\tilde{a}_A\tilde{a}_B^2 \nonumber \\
&- \left(2C_{\text{Oct} B} + 15 C_{Q B} - 3 X_{AB} C_{Q B}\right)\tilde{a}_B^3,
\end{align}
which is qualitatively different from Eq.~\eqref{EjS3} above because of
the presence of the spin-induced octupolar moments $C_{\text{Oct}A,B}$.
Comparing Eqs.~\eqref{EjS3} and \eqref{EjS3_LS}, we see that \TEOBResumS{} does not
automatically predict (through the definition Eqs.~\eqref{rc2}, \eqref{aQ} used to incorporate spin-quadratic couplings) the needed PN LO spin-cubic terms.
We have, however, checked that the coefficients entering the two expressions  are numerically sufficiently
close to lead to nearly equivalent physical predictions.
This is especially evident in the BBH case when $C_{\text{Oct}A,B}=C_{QA,B}=1$.
In this case the above equations read
\begin{align}
\left(c_{7}^{S^3_{\rm LO}}\right)^{\TEOBResumS{}} =& 
-\left(\frac{67}{4}\tilde{a}_0 + \frac{13}{4} X_{AB} \tilde{a}_{AB}\right)\tilde{a}_0^2,\\
\left(c_{7}^{S^3_{\rm LO}}\right)^{\rm L-S} =& 
-\left(17\tilde{a}_0 + 3 X_{AB} \tilde{a}_{AB}\right)\tilde{a}_0^2,
\end{align}
with a fractional difference of $1/68\approx 1.47\%$ between the first coefficients and $1/12\approx 8.3\%$ for second ones. In practice, the Hamiltonian
of \TEOBResumS{} incorporates this approximate description of
cubic-in-spin terms, as well as higher-order odd powers of the
spins due to its resummed structure. In Appendix~\ref{app:S3}
we propose possible EOB transcriptions of the full cubic-in-spin
information of Ref.~\cite{Levi:2015uxa}.

\subsection{Hamiltonian: quartic-in-spin terms}
\label{sec:HamS4}
The quartic-in-spin contribution to the PN-expanded $E_b(j)$ curve
was also computed by Levi and Steinhoff~\cite{Levi:2016ofk}. This
corresponds to a 4PN effect, i.e., it enters at order $1/j^{10}$.
We can thus slightly modify the procedure of Sec.~\ref{sec:S2} above
so to apply it also to the recovery of the spin-quartic EOS-dependent terms.
We introduce a new parameter $\delta a^4_{\rm LO}$ in the definition of $r_c^2$ that now
reads
\begin{equation}
r_c^2=r^2+\tilde{a}_{Q}^2\left(1+\frac{2}{r}\right) +\frac{\delta a^2_{\rm NLO}}{r} +\frac{\delta a_{\rm NNLO}^2}{r^2}+\frac{\delta a^4_{\rm LO}}{r^2}.
\end{equation}
We then proceed and compute the same formulas we showed before consistently keeping all the quartic-in-spin term.
The LO quartic-in-spin term, $O(1/j^{8})$ in $E_b(j)$ reads
\begin{equation}
\left(c_{8}^{S^4_{\rm LO}}\right)^{\TEOBResumS{}}= 3 \tilde{a}_Q^4+\delta a_{\rm LO}^4.
\end{equation}
The corresponding term from Ref.~\cite{Levi:2015ixa} reads
\begin{align}
\left(c_8^{S^4_{\rm LO}}\right)^{\rm L-S}=&~\frac{3}{4} \left(3C_{QA}^2+C_{\text{Hex} A}\right)\tilde{a}_A^4 \nonumber \\
&+3 \left(3C_{QA}+C_{\text{Oct} A}\right)\tilde{a}_A^3\tilde{a}_B  \nonumber \\
&+ 9 \left(C_{QA}C_{QB}+1\right) \tilde{a}_A^2 \tilde{a}_B^2 \nonumber \\
&+3 \left(3C_{QB}+C_{\text{Oct} B}\right)\tilde{a}_A\tilde{a}_B^3 \nonumber \\
&+\frac{3}{4} \left(3C_{QB}^2+C_{\text{Hex}B}\right)\tilde{a}_B^4 ,
\end{align}
where $C_\text{Oct}$ and $C_\text{Hex}$ are the spin-induced octupolar and
hexadecapolar moments quoted above. From these two equation one obtains
\begin{align}
  \label{eq:delta_a4_LO}
\delta a^4_{\rm LO} =&~ \frac{3}{4}\left(C_{\text{Hex} A}-C_{QA}^2\right)\tilde{a}_A^4 \nonumber \\
&+ 3\left(C_{\text{Oct} A}-C_{QA}\right)\tilde{a}_A^3\tilde{a}_B \nonumber \\
&+ 3\left(C_{QA}C_{QB}-1\right)\tilde{a}_A^2\tilde{a}_B^2 \nonumber \\
&+ 3\left(C_{\text{Oct} B}-C_{QB}\right)\tilde{a}_A\tilde{a}_B^3 \nonumber \\
&+\frac{3}{4}\left(C_{\text{Hex} B}-C_{QB}^2\right)\tilde{a}_B^4.
\end{align}
To our knowledge $C_Q$ and $C_\text{Oct}$ have been calculated  using
numerical approaches~\cite{Laarakkers:1997hb,Pappas:2012ns,Pappas:2012qg};
by contrast, current knowledge about $C_\text{Hex}$ relies on both the
slow-rotation approximation (if the NS dimensionelss spin is smaller than 0.3)
and on numerical calculations otherwise~\cite{Yagi:2014bxa}. All this
knowledge (notably recasted in terms of EOS quasi-universal
relations~\cite{ Yagi:2013bca,Yagi:2013awa,Yagi:2014bxa} with
the NS Love numbers~\cite{Damour:1983a,Damour:1992qi,Hinderer:2007mb,Damour:2009vw,Binnington:2009bb})
allows us to evaluate also the impact $\delta a^4_{\rm LO}$ on the BNS phasing.
Before doing so, we note that in the BH limit (when $C_Q=C_\text{Oct}=C_\text{Hex}=1$) 
$\delta a^4_{\rm LO}$ vanishes. It is remarkable that the resummed 
EOB Hamiltonian, thanks to the use of the deformed Kerr structure provided
by the EOB centrifugal radius~\cite{Damour:2014sva}, is proven to correctly  
incorporate, at the LO, the quartic-in-spin behavior. We also point out,
in passing, that the same structure is present also in the EOB Hamiltonian 
of Refs.~\cite{Barausse:2009xi,Barausse:2011ys,Taracchini:2012ig},
and thus the quartic-in-spin terms at LO are also present 
in the {\tt SEOBNRv4} corresponding EOB model~\cite{Hinderer_private}

%=============================
\subsection{Waveform and flux}
\label{sec:S2_waveform}
%=============================
Recently, Marsat and Boh\'e have also computed several terms quadratic in spin
entering the post-Newtonian waveform~\cite{Marsat_private}. Their work is
yet unpublished, but they kindly gave us access to their most recent results. 
 We report below the corresponding contributions to the factorized 
waveform amplitude, as the EOS-dependent generalization of Eqs.~(39), (43), (44) and (45) 
of Ref.~\cite{Messina:2018ghh}.
\begin{widetext}
\begin{align}
\label{eq:rho22}
\rho_{22}^{\rm SS, LO}=&~\frac{1}{2}\tilde{a}_{Q}^2 x^2,\\
\rho_{22}^{\rm SS, NLO}=&\left\{ -\frac{187}{252}\tilde{a}_0^2+\left(\frac{1}{7}+\frac{27}{56}\nu\right)\tilde{a}_{Q}^2+\left(\frac{19}{252}-\frac{5}{18}\nu\right)\tilde{a}_{AB}^2+ X_{AB}\left[\frac{2}{9}\tilde{a}_0\tilde{a}_{AB} +\frac{55}{84}(C_{QA}\tilde{a}_A^2-C_{QB}\tilde{a}_B^2)\right]\right\} x^3,\\
\tilde{f}_{21}^{\rm SS,LO}=&\left[ -\frac{19}{8}\tilde{a}_0\tilde{a}_{AB}-(C_{QA}\tilde{a}_A^2-C_{QB}\tilde{a}_B^2)
+X_{AB}\left(-\tilde{a}_0^2+\frac{3}{2}\tilde{a}_{Q}^2-\frac{1}{8}\tilde{a}_{AB}^2\right)\right] x^2,\\
\tilde{f}_{31}^{\rm SS,LO}=&\left[-4(C_{QA}\tilde{a}_A^2-C_{QB}\tilde{a}_B^2)+\frac{3}{2}X_{AB}\tilde{a}_{Q}^2\right]x^2,\\
\label{eq:f33}
\tilde{f}_{33}^{\rm SS,LO}=&~\frac{3}{2}X_{AB}\tilde{a}_{Q}^2x^2.
\end{align}
\end{widetext}
For this work, all these new terms (due to Marsat and Boh\'e) are incorporated
in the flux and waveform of \TEOBResumS{}.

Let us finally comment about the cubic-in-spin terms, that, at leading
order, contribute to both the $\ell=m=2$ and to the $\ell=2$, $m=1$
quadrupolar modes. The corresponding contribution to the flux was obtained
by S.~Marsat in Ref.~\cite{Marsat:2014xea}. In Ref.~\cite{Messina:2018ghh}
this information (though restricted to the BBH case) was incorporated
in the EOB waveform. Although the results of this paper are obtained
by {\it omitting} such LO spin-cube contribution, let us write here
the full terms entering $\rho_{22}^{S}$ and $\tilde{f}_{21}^S$, that reduce
to part of Eqs.~(39) and~(43) of Ref.~\cite{Messina:2018ghh} in the black hole limit
$C_{QA}=C_{QB}=C_{\text{Oct}A}=C_{\text{Oct}B}=1$.
\begin{align}
\rho_{22}^{S^3}=&
~\biggl\lbrace \left(\frac{19}{12}C_{QA}-C_{\text{Oct}A}-\frac{1}{4}C_{QA}X_{AB}\right)\tilde{a}_A^3\nonumber\\
&
+\biggl[\frac{19}{6}-\frac{17}{12}C_{QA}-\left(\frac{1}{2}-\frac{1}{4}C_{QA}\right)X_{AB}\biggr]\tilde{a}_A^2\tilde{a}_B\nonumber\\
&
+\biggl[\frac{19}{6}-\frac{17}{12}C_{QB}+\left(\frac{1}{2}-\frac{1}{4}C_{QB}\right)X_{AB}\biggr]\tilde{a}_A \tilde{a}_B^2\nonumber\\
&
+\left(\frac{19}{12}C_{QB}-C_{\text{Oct}B}+\frac{1}{4}C_{QB}X_{AB}\right)\tilde{a}_B^3\biggr\rbrace x^{7/2},\\
\tilde{f}_{21}^{S^3}=&
~\left(\frac{3}{2}\tilde{a}_0^2-\frac{3}{4}\tilde{a}_Q^2\right)\tilde{a}_{AB}~ x^{5/2}.
\end{align}

%====================================================
\section{Post-Newtonian phasing description}
\label{sec:PN}
%====================================================

\subsection{Reminder on the the TaylorF2 phasing approximant}

In the previous section, we have extensively discussed the spin-quadratic
(and spin-quartic) contributions in both the Hamiltonian and
waveform/flux of \TEOBResumS{}. Inspecting the expressions for,
e.g., $(\delta a^2_{\rm NLO},\delta a^2_{\rm NNLO})$ one sees that
there are several terms that involve $(C_{QA},C_{QB})$ and thus take
into account the effect due to the spin-induced quadrupole moments
both in the dynamics and in the radiation (cf. Eqs.~\eqref{eq:rho22}-\eqref{eq:f33}).
In this section, we move to the PN-based equivalents of these effects within
the TaylorF2 phasing approximant~\cite{Damour:2000zb}.
Although our final goal is to compare the effect of the spin-induced
quadrupole moment in TaylorF2 and in \TEOBResumS{}, here
we aim at being as general as possible. So, for completeness we
collect all currently available spin-dependent analytical
information that allows us to push the complete spin sector
of the TaylorF2 approximant up to 4PN accuracy. This means
considering linear and quadratic-in-spin effects that also involve 
tail terms. 

Given the energy flux $\mathcal{F} (v)$ and energy $E(v)$ of a binary 
on circular orbits (expressed as functions of $v \equiv (M \Omega)^{\frac13}$,
where $\Omega$ is the orbital frequency), the phase of the Fourier transform of the signal is obtained by
the following integral (see, e.g., Eq.~(3.5) of Ref.~\cite{Damour:2000zb})
\begin{equation}
\label{eq:getTF2}
\Psi(f)= 2\pi f t_{\rm ref}-\phi_{\rm ref}+2\int_{v_f}^{v_{\rm ref}}(v_f^3-v^3)\frac{E'(v)}{\mathcal{F}(v)}dv,
\end{equation}
which assumes the validity of the stationary phase approximation.
In short, when this approximation holds, the phase
of the time-domain Fourier transform is the Legendre transform of
the quadrupolar time domain phase $\phi (t)$, that is given by
\begin{equation}
\label{eq:Psif}
\Psi(f) = 2\pi f t_{f}-\phi(t_{f})-\pi/4,
\end{equation}
where $t_{f}$ is the solution of the equation  $2\pi f = [d\phi(t)/dt]_{t=t_f}$. Using in Eq.~\eqref{eq:getTF2}
the PN-expanded expression for $E'(v)/\mathcal{F}(v)$ defines the TaylorF2
phasing approximant. TaylorF2 is typically used at 3.5PN accuracy for the orbital 
and spin-orbit part, while the spin-spin part is limited to 3PN order. It was used in this
form for GW data-analysis purposes (see, e.g.,~\cite{Abbott:2018wiz,TheLIGOScientific:2017qsa}).
The complete extension of the approximant at 4PN is currently not possible since the calculation
of the $\nu$-dependent part of the energy fluxes is currently incomplete. However, there are higher-order
terms in TaylorF2, those involving the tail terms, that are analytically known. For example, Ref.~\cite{Messina:2017yjg} showed how the 4.5PN-accurate
term of the energy flux, that is a pure tail term, can be obtained exactly by PN-expanding
the EOB energy flux. Applying the same procedure, one can have access to the 3.5PN-accurate,
LO, spin-spin tail term as well as to the 4PN-accurate, NLO, spin-orbit tail term.
The spin-spin and spin-orbit tail terms in the flux (and TaylorF2) are presented here
for the first time. After the integration of Eq.~\eqref{eq:getTF2} the 4PN-accurate
spin-dependent part of the phasing reads
\begin{align}
\label{eq:Psi_spa}
\Psi_{\rm 4PN, spin}^{\rm F2}(f)=&~2\pi f t_c-\varphi_c-\frac{\pi}{4}\nonumber\\
&+\frac{3}{128\nu}(\pi f M)^{-5/3}\sum_{i=0}^{8}\varphi_i(\pi f M)^{i/3}.
\end{align}
As mentioned above, the orbital (spin-independent) part has the same structure,
but the 4PN term is currently incomplete, so we omit its discussion here.
Following the procedure of Ref.~\cite{Messina:2017yjg}, we construct the PN-expanded
total energy flux starting from the EOB-resummed prescription~\cite{Damour:2008gu}
\begin{equation}
\label{eq:Flux}
{\cal F}=\sum_{\ell=2}^{\infty}\sum_{\ell =-m}^{m} F^{\rm Newt}_{\ell m}\hat{F}_{\ell m}
\end{equation}
using the orbital dynamical information at the consistent
PN order~\cite{Bernard:2016wrg,Damour:2016abl,Damour:2015isa,Damour:2014jta,Blanchet:2013haa},
the spin information given in Ref.~\cite{Bohe:2015ana} and the new spin waveform results 
computed by Marsat (Eqs. (18)-(22) from Sec.~\ref{sec:S2_waveform},
\cite{Messina:2017yjg,Messina:2018ghh,Cotesta:2018fcv}).
The relation between the dynamics, the EOB residual relativistic
amplitudes (which can be derived from the PN waveforms) and the flux is
given in Ref. \cite{Damour:2008gu}, and we reproduce it here for completeness:
\begin{equation}
\label{eq:hF}
\hat{F}_{\ell m} = \left(S^{(\epsilon)}_{\rm eff}\right)^{2}|T_{\ell m}|^{2}(\rho_{\ell m})^{2\ell}.
\end{equation} 
In this equation, $S^{(\epsilon)}_{\rm eff}$ is the effective source, that is the effective
EOB energy along circular orbits $\hat{E}_{\rm eff}(x)\equiv E_{\rm eff}/\mu$ when $\epsilon=0$ ($\ell+m$=even)
or the Newton-normalized orbital angular  momentum when $\epsilon=1$ ($\ell+m$=odd). 

\subsection{Extracting tail effects from the EOB  resummed tail factor $T_{\ell m}$}

Of crucial importance for our present purpose is the (complex) tail factor $T_{\ell m}$ that resums an infinite number of leading logarithms (see Refs.~\cite{Damour:2008gu,Faye:2014fra}).  This factor automatically incorporates tail effects that can
be extracted from it and added to the lower-order PN results.

Expanding the formula \eqref{eq:hF} multipole by multipole and then summing all
the contributions up to $\ell=4$, one  obtains the following expression for the 3.5PN spin-quadratic tail term
in the flux
\begin{equation}
\hat{{\cal F}}^{\rm SS, tail}_{\rm 3.5PN}=\left(8 \tilde{a}_Q^2 + \frac{1}{8}\tilde{a}_{AB}^2\right)\pi ~x^{7/2},
\end{equation}
which reduces to Eq.~(26) of Ref.~\cite{Messina:2017yjg} in the BH case, when $\tilde{a}^2_Q=\tilde{a}^2_0$.
Adding this new piece to Eq. (4.14) of Ref.~\cite{Bohe:2015ana}, one obtains a full 3.5PN flux
that is used, together with the energy given by Eq.~(3.33) of the same reference,
to compute the 3.5PN accurate spin-spin tail term at NLO (entering the $\varphi_7$ coefficient 
in Eq.~\eqref{eq:Psi_spa}, as detailed below) by solving the integral given by Eq.~\eqref{eq:getTF2}.

This resummed tail expansion procedure can be applied also to the spin-orbit analogue of the flux.
As we did previously, the fact that the EOB-resummed tail amplitude $T_{\ell m}$ contains 
and infinite amount of PN information when expanded, using consistently the 
$\rho_{\ell m}$ and $\tilde{f}_{\ell m}$ information computed from Eqs.~\eqref{eq:rho22}-\eqref{eq:f33} and the
point-mass ones from~\cite{Messina:2018ghh,Cotesta:2018fcv},
we can compute again Eq.~\eqref{eq:hF}, but this time, for what concerns the dynamics, we add to the
orbital information of Refs.~\cite{Damour:2014jta,Damour:2015isa,Damour:2016abl,Bernard:2016wrg}
the spin-orbit one of Refs.~\cite{Marsat:2014xea,Bohe:2013cla}. The spinning angular momentum at
NLO in the spin-orbit coupling is given by Eq. (3.12) of Ref.~\cite{Marsat:2013caa}.
This time we use the spin residual relativistic waveform amplitudes up to $\ell=4$, and the purely
orbital ones from $\ell=5$ to $\ell=7$, truncating at the right PN order being careful to account
for the relative order of the Newtonian prefactors in the process (see Appendix of~\cite{Messina:2018ghh}).
Like in Ref.~\cite{Messina:2017yjg}, the $m={\rm even}$ flux information of the $\ell=7$ multipoles
is out of one PN order with respect to the result we are searching for, so can be neglected in this
computation. The new result obtained this way yields~\footnote{By PN consistency,
this procedure yields lower PN spin-orbit terms that are well
known in literature. Note also that the test-particle limit of \eqref{eq:5PNtailSOF} 
(in which $\tilde{a}_0=\tilde{a}_{AB}X_{AB}=\tilde{a}_A$) agrees with the result of Ref.~\cite{Fujita:2014eta}, namely $\frac{23605}{144}\tilde{a}_A\pi x^5$.}
\begin{align}
\label{eq:5PNtailSOF}
&\mathcal{F}_{\rm 5PN}^{\rm SO, tail}=\biggl[\biggl(\frac{220103}{1512}+\frac{8421757}{72576}\nu-\frac{9491453}{18144}\nu^2\biggr)\tilde{a}_0\nonumber\\
&+\biggl(\frac{55499}{3024}+\frac{1149163}{72576}\nu-\frac{4993897}{36288}\nu^2\biggr)X_{AB}\tilde{a}_{AB}\biggr]\pi x^5,
\end{align}
and gives access to the new 4PN spin-orbit terms in the $\varphi_8$ coefficient in the PN-SPA phase.

\subsection{Final 4PN-accurate TaylorF2 phasing coefficients}

The complete calculation, at 4PN-accuracy, gives
\begin{widetext}
\begin{align}
\label{eq:spintf2}
\varphi_3^{\rm SO}=&~\frac{94}{3}\tilde{a}_0 +\frac{19}{3}X_{AB}\tilde{a}_{AB},\\
\varphi_4^{\rm SS}=&-50 \tilde{a}_Q^2 - \frac{5}{8}\tilde{a}_{AB}^2,\\
\varphi_5^{\rm SO}=&-\left[1+\log\left(\pi f M\right)\right]\left[\left(\frac{554345}{2268}+\frac{55}{9}\nu\right)\tilde{a}_0+\left(\frac{6380}{81}+\frac{85}{9}\nu\right) X_{AB}\tilde{a}_{AB}\right],\\
\varphi_6^{\rm SS;~\rm SO_{\rm tail}}=&~\pi\left(\frac{1880}{3}\tilde{a}_0+130X_{AB}\tilde{a}_{AB}\right) +\left(\frac{15635}{21}+120\nu\right) \tilde{a}_Q^2 - \frac{5570}{9}\tilde{a}_0^2 \nonumber\\
&-\left(\frac{40795}{2016}+\frac{1255}{36}\nu\right)\tilde{a}_{AB}^2+X_{AB}\left[-\frac{250}{9}\tilde{a}_0\tilde{a}_{AB}+\frac{2215}{12}\left(C_{QA}\tilde{a}_A^2-C_{QB}\tilde{a}_B^2\right)\right],\\
\varphi_7^{\rm SS_{tail};~SO}=&-\pi\left(400 \tilde{a}_Q^2 + \frac{15}{2}\tilde{a}_{AB}^2\right) + \biggl(-\frac{8980424995}{1524096}+\frac{6586595}{1512}\nu-\frac{305}{72}\nu^2\biggr)\tilde{a}_0 \nonumber\\
&+\left(-\frac{7189233785}{3048192}+\frac{458555}{6048}\nu-\frac{5345}{144}\nu^2\right)X_{AB}\tilde{a}_{AB},\\
\varphi_8^{\rm SO_{tail}}=&~\pi[1-\log\left(\pi f M\right)]\left[\left(\frac{2388425}{2268}-\frac{9925}{27}\nu\right)\tilde{a}_0+\left(\frac{1538855}{4536}-\frac{19655}{756}\nu\right)X_{AB}\tilde{a}_{AB}\right],
\end{align}
\end{widetext}
where we have explicitly emphasized in the definition of each term its spin-orbit, spin-spin or spin-tail character.
Note that $\varphi_6$ and $\varphi_7$ receive contributions from both tail and non-tail terms.
The $(\varphi_7^{\rm SS_{\rm tail}},\varphi_8^{\rm SO_{\rm tail}})$ terms are computed here from the first time. 

\subsection{Isolating the EOS-dependent quadrupole-monopole terms}

From the result above we can finally isolate the EOS-dependent quadrupole-monopole terms
(i.e., those proportional to $C_{Qi}$). These terms are the main focus of the present paper.
Multiplying by the Newtonian prefactor (see also~\cite{Nagar:2018zoe})
one obtains, at 3.5PN order, 
\begin{equation}
\Psi_{\rm SS}^{\rm QM} = \Psi^{\rm QM,LO}_{\rm SS} + \Psi^{\rm QM, NLO}_{\rm SS} + \Psi^{\rm QM, tail}_{\rm SS},
\end{equation}
which can be explicitly written as
\begin{align}
\label{eq:PsiPNqm}
\Psi^{\rm QM,LO}_{\rm SS}=& -\dfrac{75}{64\nu}\left(\tilde{a}_{A}^{2}C_{QA}+\tilde{a}_{B}^{2}C_{QB}\right)\left(\dfrac{M \omega}{2}\right)^{-1/3},\\
\label{eq:PsiPNqmNLO}
\Psi^{\rm QM, NLO}_{\rm SS}=&~\frac{1}{\nu}\biggl[\left(\frac{45}{16}\nu+\frac{15635}{896}\right)(C_{QA}\tilde{a}_A^2+C_{QB}\tilde{a}_B^2)\nonumber\\
  &+\frac{2215}{512}X_{AB}(C_{QA}\tilde{a}_A^2-C_{QB}\tilde{a}_B^2)\biggr]\left(\frac{M \omega}{2}\right)^{1/3},\\
\label{eq:PsiPNqmNNLO}
\Psi^{\rm QM, tail}_{\rm SS}=& -\dfrac{75}{8\nu}\pi\left(\tilde{a}_{A}^{2}C_{QA}+\tilde{a}_{B}^{2}C_{QB}\right)\left(\dfrac{M \omega}{2}\right)^{2/3}
\end{align}
where we replaced the circularized quadrupolar gravitational wave frequency $Mf$ by $M \omega=2\pi M f$.
The LO (2PN) term has been derived from the SPA phase originally computed by
Poisson~\cite{Poisson:1997ha}.
%=============================================
% Table with the details of the configurations
%=============================================
\begin{table*}[t]
\begin{ruledtabular}
\begin{tabular}{lccccccccccccc}
No. & EOS  & $M_A$  & $M_B$ & $q\equiv M_A/M_B$ & $\chi_A$ & $\chi_B$ & $\Lambda_A$ & $\Lambda_B$ & $C_{QA}$ & $C_{QB}$ & $M\omega_{\rm mrg}$ & $\Delta^{\rm EOB} \phi$ \\ 
\hline
\hline
{\tt SLy-q1-sA01-sB01}       & Sly  & 1.35   & 1.35  &  1  	  & 0.1  & 0.1  & 389.96 & 389.96 & 5.48 & 5.48 & 0.1344     & $-2.04$ \\
{\tt SLy-q1-sA005-sB005}     & Sly  & 1.35   & 1.35  &  1 	  & 0.05 & 0.05 & 389.96 & 389.96 & 5.48 & 5.48 & 0.13446    & $-0.51$ \\
{\tt SLy-q1.2-sA005-sA008}   & Sly  & 1.6573 & 1.354 &  1.224      & 0.05 & 0.08 & 382.7  & 1312.1 & 5.45 & 7.99 & 0.12155    & $-1.04$ \\  
{\tt Ms1b-q1-sA01-sB01}      & Ms1b & 1.35   & 1.35  &  1 	  & 0.1  & 0.1  & 1545   & 1545   & 8.40 & 8.40 & 0.10616    & $-3.06$ \\  
{\tt Ms1b-q1-sA005-SB005}    & Ms1b & 1.35   & 1.35  &  1          & 0.05 & 0.05 & 1545   & 1545   & 8.40 & 8.40 & 0.10616    & $-0.76$ \\  
{\tt H4-q1.25-sA005-sB008}   & H4   & 1.91   & 1.528 &  1.25       & 0.05 & 0.08 & 499.6  & 1986   & 5.92 & 9.06 & 0.11508    & $-1.08$ \\
\end{tabular}
\end{ruledtabular}
\caption{
  \label{tab:RunsTable} BNS configurations used in this paper. From left to right the columns report:
  the name of the configurations; the individual masses; the mass ratio; the individual spins, tidal
  parameters and spin-induced quadrupole moments, that are obtained with the universal relations
  of~\cite{Yagi:2013bca,Yagi:2013awa}.
  Then, $M\omega_{\rm mrg}$ denotes the dimensionless GW frequency at the EOB BNS merger, conventionally defined
  as the peak of the $\ell=m=2$ waveform amplitude. The last column lists the accumulated phase from
  10~Hz to BNS merger due to the presence of the self-spin effects. For obtaining these numbers only
  the NLO self-spin terms, both in waveform and Hamiltonian, were included in \TEOBResumS{}.}
\end{table*}
%========================================================================================================================================
%==========================
% First cases: NLO only
%==========================
\begin{figure*}[t]
	\center
	\includegraphics[width=0.48\textwidth]{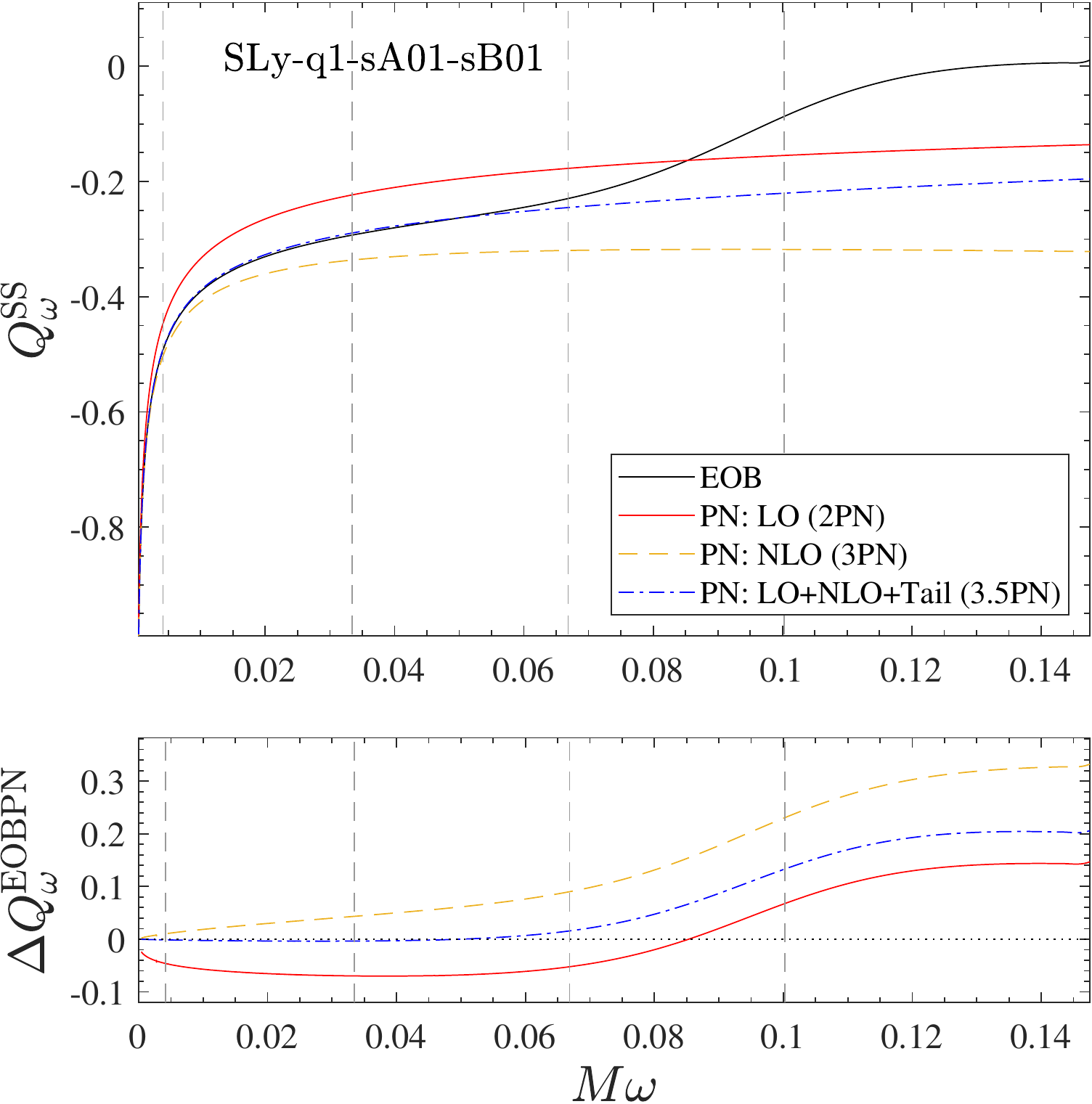}
        \hspace{5mm}
        \includegraphics[width=0.48\textwidth]{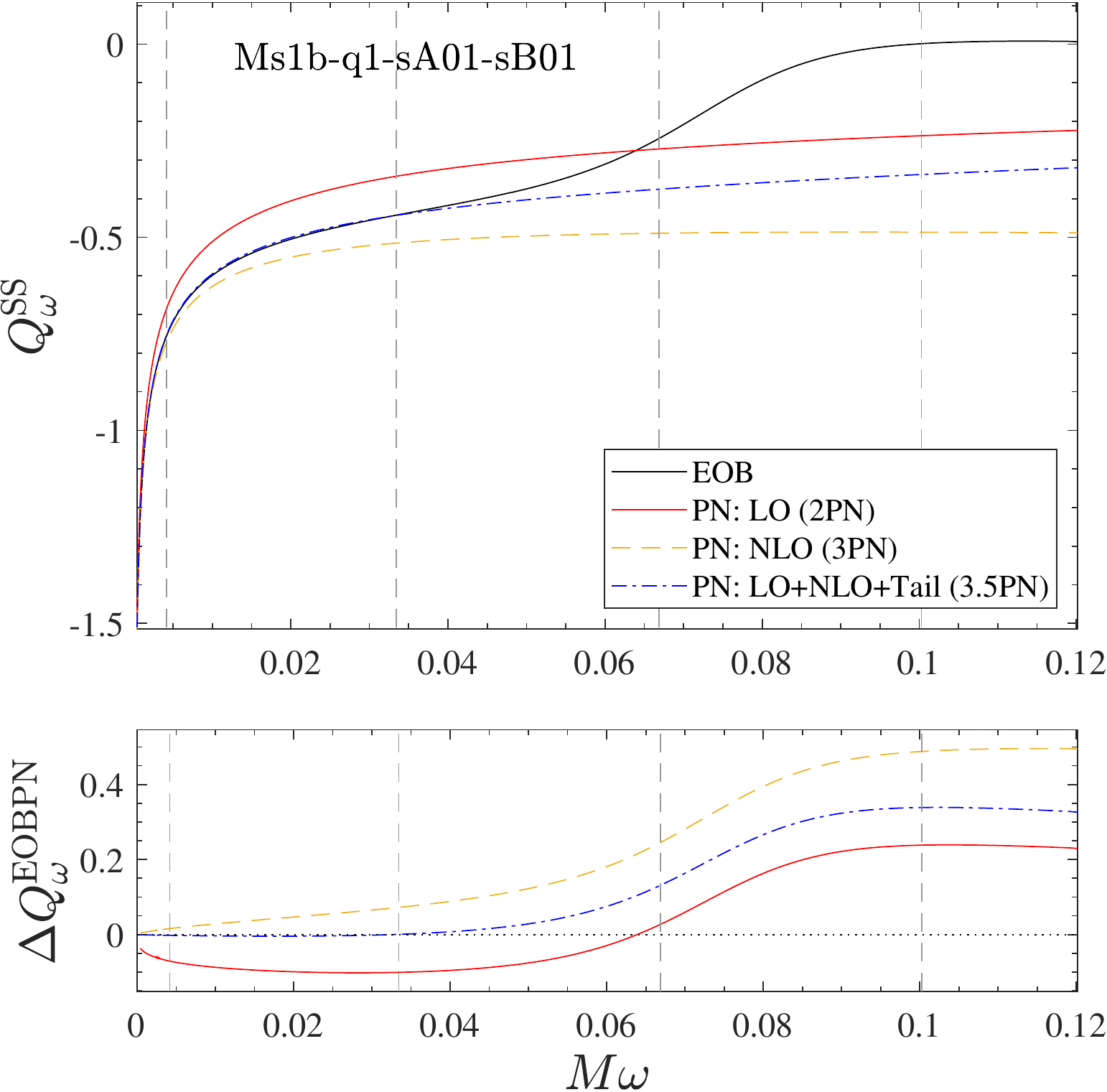}  
	\caption
	{	
		\label{fig:compare_nlo} 
		Left panel: Sly EOS, $M_A=M_B=1.35$, $\chi_A=\chi_B=0.1$ and $C_{QA}=C_{QB}=5.48$.
                Right panel: Ms1B EOS, $M_A=M_B=1.35$, $\chi_A=\chi_B=0.1$ and $C_{QA}=C_{QB}=8.40$.
		The grey vertical lines correspond to 50, 400, 800 and 1200 Hz respectively.
                The additional $Q_\omega^{\rm SS_{QM,tail}}$ term in TaylorF2 is crucial to get an
                excellent agreement between the PN-expanded and EOB phasing for most of the
                inspiral. 
	}
\end{figure*}

Le us now concentrate on the gauge-invariant description of the phasing of the case of our interest,
i.e., Eqs \eqref{eq:PsiPNqm}, \eqref{eq:PsiPNqmNLO} and \eqref{eq:PsiPNqmNNLO}.
Following previous practice (see Ref.~\cite{Nagar:2018zoe} and references therein)
we do so by the $Q_{\omega}$ function that is defined, for any Fourier-domain phase,
as
\begin{equation}
\label{getQomg}
\omega^{2}\dfrac{d^{2}\Psi(\omega)}{d\omega^{2}}=Q_{\omega}(\omega),
\end{equation}
where we identify the time domain and frequency domain circular 
frequencies, i.e., $\omega_{f}=\omega(t)$.
The integral of $Q_{\omega}$ per logarithmic frequency yields the phasing
accumulated by the evolution on a given frequency interval $(\omega_L,\omega_R)$.
The quadrupole-monopole contribution to the PN-expanded $Q_\omega$ 
we are interested in here is given by the following three terms
\begin{equation}
Q_\omega^{\rm SS_{QM}} = Q_\omega^{{\rm SS}_{\rm QM,LO}}+ Q_\omega^{{\rm SS}_{\rm QM,NLO}} +  Q_\omega^{{\rm SS}_{\rm QM,tail}}
\end{equation}
that explicitly read
\begin{align}
\label{eq:QomgMQpn}
Q_{\omega}^{{\rm SS}_{\rm QM,LO}}=&-\dfrac{25}{48\nu}\left(\tilde{a}_{A}^{2}C_{QA}+\tilde{a}_{B}^{2}C_{QB}\right)\left(\dfrac{\omega}{2}\right)^{-1/3},\\
\label{eq:QomgMQpnNLO}
Q_{\omega}^{\rm SS_{QM, NLO}}=&-\frac{1}{\nu}\biggl[\left(\frac{5}{8}\nu+\frac{15635}{4032}\right)(C_{QA}\tilde{a}_A^2+C_{QB}\tilde{a}_B^2)\nonumber\\
  &+\frac{2215}{2304}X_{AB}(C_{QA}\tilde{a}_A^2-C_{QB}\tilde{a}_B^2)\biggr]\left(\frac{\omega}{2}\right)^{1/3},\\
Q_{\omega}^{\rm SS_{QM, tail}}=&~\dfrac{25}{12\nu}\pi\left(\tilde{a}_{A}^{2}C_{QA}+\tilde{a}_{B}^{2}C_{QB}\right)\left(\dfrac{\omega}{2}\right)^{2/3}.
\end{align}
The aim of the next section will be to investigate how this function compares with the analogous
quantity obtained from \TEOBResumS{} with all the spin-dependent information detailed in the
previous section.

%=====================================
\section{Results: gauge-invariant phasing comparisons of the EOS-dependent self-spin effects}
\label{sec:results}
%====================================
%=====================
% High spin comparison
%=====================
\begin{figure}[t]
	\center
	\includegraphics[width=0.48\textwidth]{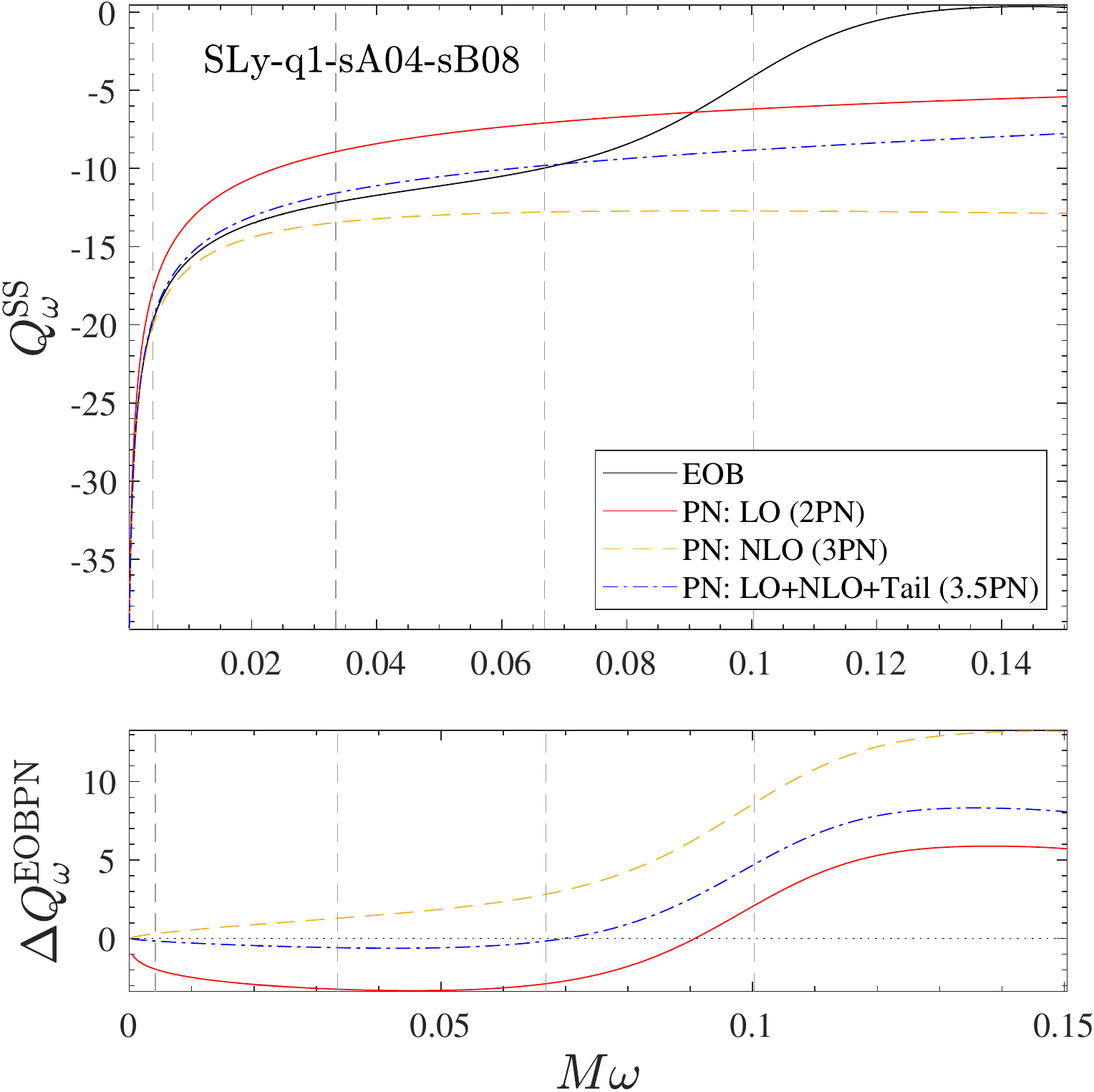}
	\caption{\label{fig:Qomg_highspin}Closeness of the tail-completed TaylorF2 description of the phasing
          to the EOB one, even when rather high values of the individual spins are considered.}
\end{figure}
Reference~\cite{Nagar:2018zoe} presented a preliminary comparison between the
various PN truncations of the $Q_\omega^{\rm SS_{QM}}$ written above and the corresponding
quantity computed using from the time-domain waveform generated by \TEOBResumS{}
including only the self-spin information at LO in both the Hamiltonian and waveform/flux,
see Fig.~14 there. The main outcome of this preliminary comparison was to show,
for an illustrative BNS configuration, the consistency between the PN and EOB
descriptions, especially at low frequencies, with the latter being slightly
more phase-accelerating than the former. In this respect, Refs.~\cite{Nagar:2018zoe,Dietrich:2018uni}
showed the existence of a nonnegligible difference with respect to the TaylorF2 phasing
with NLO (i.e., 3PN) self-spin effects. However, Ref.~\cite{Dietrich:2018uni},
see Sec.~VI there, stressed that a more definitive
assessment of the EOB/PN performances would need
 the incorporation of the NLO information in \TEOBResumS{}.
We shall do so here, closely following what was done in Ref.~\cite{Nagar:2018zoe}.
To start with, we work at NLO in the self-spin within \TEOBResumS{},
adding the corresponding terms to both the Hamiltonian and the multipolar
waveform amplitude (and flux). For definiteness, we consider a few BNS
configurations, that we list in Table~\ref{tab:RunsTable}, ranging
from stiffer to softer EOS. Similarly, we mainly explore values of
the spins that are compatible with those expected for BNSs. However,
to stretch the limits of the model, we also consider a fast-spinning
configuration, with $\chi_A=0.8$ and $\chi_B=0.4$.
We note that, although such a configuration is unlikely to exist in a realistic
binary system, these spin magnitudes values were considered in the
parameter estimation of GW170817 when considering high-spin priors
analyses~\cite{TheLIGOScientific:2017qsa,Abbott:2018wiz}.

%================================
% Runtime for an illustrative run
%================================
\begin{table}[t]
\caption{
  \label{tab:codeprest}
  Configuration {\tt SLy-q1.2-sA005-sA008} in Table~\ref{tab:RunsTable}.
  Indicative time, $\tau$, obtained with the {\tt v0.1} version of the
  public implementation of \TEOBResumS{}, needed for obtaining the complete
  waveform, from 5Hz up to merger.
  To ease the computation (and to reduce the high-frequency noise due to
  the oversampling of the inspiral part), the waveform is computed joining
  together three pieces (starting at $r_{\rm max}$ and ending at $r_{\rm min}$)
  obtained with different sampling rates $\Delta t$. Note that the value
  of $\tau$ also takes into account the time needed to actually
  write the data on disk. Runs on an Intel Core i5-8250 (1.6GHz) and 8GB RAM.
  The code was compiled with the g++ GNU compiler using O3 optimization.}
\begin{center}
\begin{ruledtabular}  
\begin{tabular}{cccccc}
  $f_0$ [Hz] &$r_{\rm max}$  & $r_{\rm min}$  & $\Delta t^{-1}$ [Hz] & $\Delta t/M$ & $\tau$ [sec] \\
  \hline                         
  5    & 264.11 & 80     &  100    &    674.2                     & 102.177 \\
  20   & 104.81 & 8      &  10000  &    6.742                     & 1.622 \\
  200  & 22.58  & merger &  100000 &    0.674                     & 1.4832\\
\end{tabular}
\end{ruledtabular}
\end{center}
\end{table}
%==============================================================================
The   $Q_{\omega}(\omega)$ is computed (from the time-domain phasing)
in the same way as briefly described in Ref.~\cite{Nagar:2018zoe},
though here we pushed the lower frequency limit down to 5Hz, so as
to unambiguously identify the frequency region were the EOB and PN curve converge together.
The code we used to do so is \TEOBResumS{} {\tt v0.1} that improves over
{\tt v0.0} (see Ref.~\cite{Nagar:2018zoe}) because of the presence of the
nonlinear spin terms discussed here\footnote{Note that {\tt v0.1} also implements
  by default the EOS-dependent quartic-in-spin terms of Sec.~\ref{sec:HamS4}.}
The same terms are also implemented in {\tt v1.0} which additionally contains the
updated tidal model of Ref.~\cite{Akcay:2018yyh} and the post-adiabatic
approximation to efficiently compute long inspirals~\cite{Nagar:2018gnk}
(see Appendix~\ref{sec:PA} for additional details).
With \TEOBResumS{} {\tt v0.1} is is easy to compute such a long waveform
with reasonable efficiency. In practice, it is convenient to join together
the waveforms computed on three different frequency intervals.
The intervals are chosen in a way that
the two pieces overlap on a common frequency interval. Table~\ref{tab:codeprest}
illustrates our choices for one specific configuration, {\tt SLy-q1.2-sA005-sA008}
in Table~\ref{tab:RunsTable}, and lists: the different running times $\tau$ (including
the time needed to write the file on disk); the sampling $\Delta t/M$; and the various
intervals (in radius) where the EOB dynamics is evolved.
As explained in Ref.~\cite{Nagar:2018zoe}, in order to explore the low-frequency
regime one has to avoid the time-domain oversampling of the waveform that naturally
occurs from the ODE solver. To remove this, the raw time-domain waveform phase is
additionally downsampled and its derivatives smoothed in order to get a clean and
nonoscillatory $Q_\omega$ function. The procedure is tedious, but straightforward
and it is done separately on different frequency intervals, with the final results
eventually joined together. To isolate the, $C_{Qi}$-dependent only, $Q_\omega^{\rm SS}$
contribution within \TEOBResumS{} we perform, for each configuration,
two different runs, one with $C_{Qi}\neq0$ another one with $C_{Qi}= 0$.
In both cases we compute the time-domain $Q_\omega$ and finally calculate
\begin{equation}
Q_\omega^{\rm TEOBResumS,SS} \!\!= Q_\omega^{\rm TEOBResumS_{C_{Qi}\neq0}}\!\!\!-Q_\omega^{\rm TEOBResumS_{C_{Qi}=0}}.
\end{equation}
Illustrative results are shown in Fig.~\ref{fig:compare_nlo} for the two
configurations {\tt SLy-q1-sA01-sB01} (left panel) and {\tt Ms1b-q1-sA01-sB01} (right panel).
Each panel is separated into two subpanels: the top part reports $Q_\omega^{\rm SS}$, with the
EOB and the three different PN truncations; the bottom panel reports the differences
$\Delta Q_{\omega}^{\rm EOBPN}\equiv Q_{\omega}^{\rm SS_{EOB}}-Q_{\omega}^{\rm SS_{PN}}$. 
To orient the reader, the vertical lines superposed to the plot correspond to 50, 400, 800 and 1200~Hz.
As mentioned in Ref.~\cite{Nagar:2018zoe} the comparison between the time-domain
EOB $Q_\omega$ and the frequency domain PN-expanded $Q_\omega$ is meaningful
as long as the SPA holds. In other words, this is true until the adiabatic parameter given
by $1/Q_\omega$ is small enough. We will briefly comment about this at the end of
the section. 

The main conclusions we draw from figure Fig.~\ref{fig:compare_nlo} are:
 (i) the EOB description of the self-spin effects at NLO is more phase-accelerating
than the LO one (both PN or EOB, cf. Fig.~14 of Ref.~\cite{Nagar:2018zoe}); (ii) it
is however {\it less phase-accelerating} than the standard TaylorF2 NLO one. This seems  to
corroborate the suggestion made in Sec.~VI of Ref.~\cite{Dietrich:2018uni} that
part of the phasing accumulated by this approximant is due to its PN-nature. Since, a priori,
tidal effects might be degenerate with  self-spin effects (since they both accelerate the phasing),
 the use of the 3PN Taylor-expanded approximant may introduce biases in
the measurement of the tidal parameters. This will deserve further investigations in the
future; (iii) on the other hand, our comparisons show that the TaylorF2 phasing augmented with the tail
factor is fully consistent with the EOB-resummed description up to frequencies $\sim 600$~Hz. 
We have checked that the importance of the self-spin tail term in reconciling the TaylorF2
with the EOB phasing description remains essentially the same when changing the BNS model, though
it slightly deteriorates when the individual spins are increased.
For example Fig.~\ref{fig:Qomg_highspin} refers to the {\tt SLy-q1-sA04-sB08} configuration,
with $\chi_A=0.4$ and $\chi_B=0.8$, which illustrates the ability of the tail-completed
TaylorF2 approximant to reasonably agree with the EOB phasing even in difficult corners
of the parameter space.
%==========================
% Going to NNLO
%==========================
\begin{figure}[t]
	\center
	\includegraphics[width=0.48\textwidth]{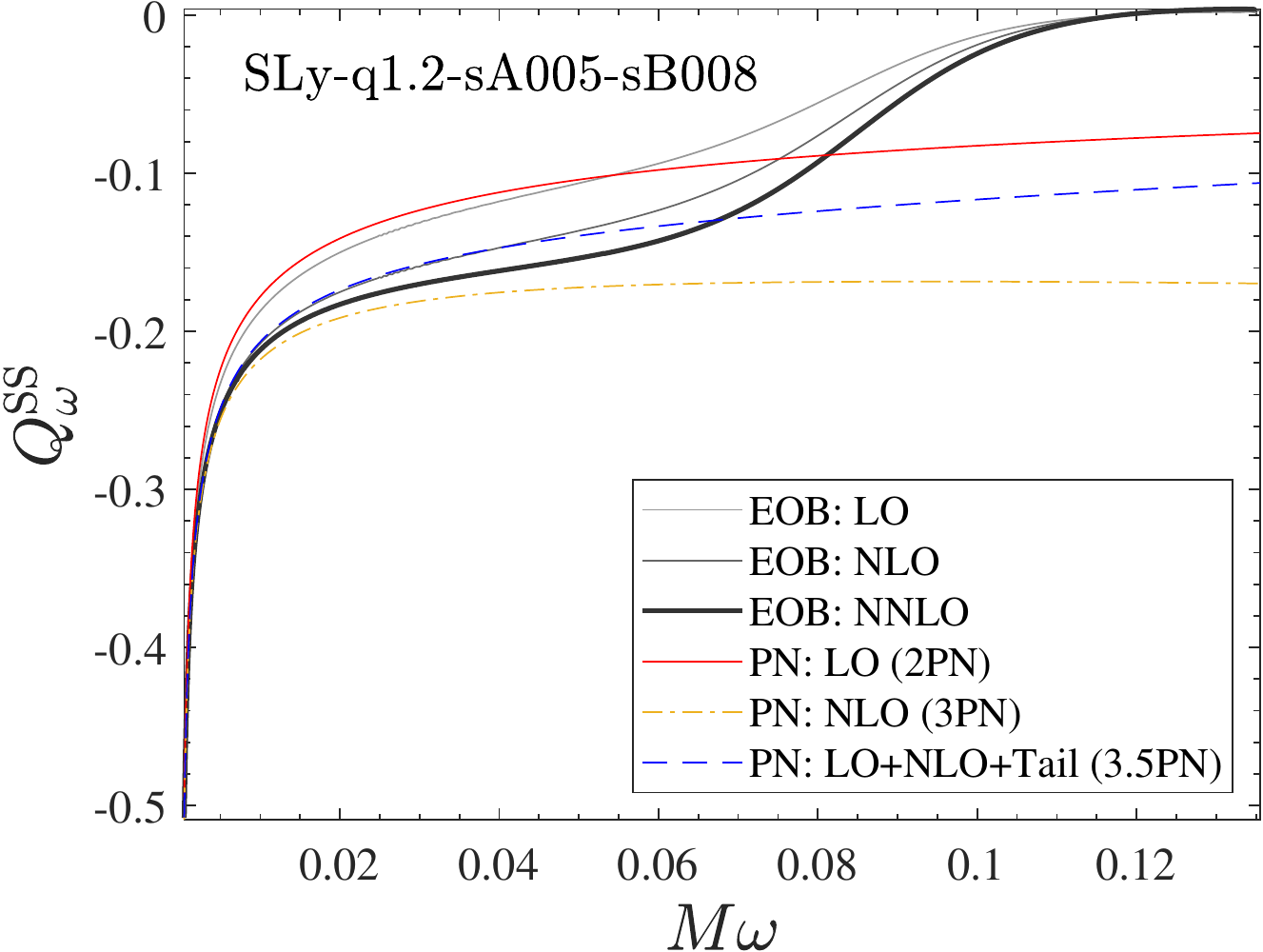}
	\caption{\label{fig:Qomg_nnlo}The effect of the NNLO self-spin term incorporated in the EOB
          Hamiltonian for one of the configurations of Table~\ref{tab:RunsTable}. Although the NNLO
          term results in an acceleration of the inspiral with respect to the NLO model, the curve
          is still above the NLO PN-expanded TaylorF2 one.}
\end{figure}
%----------------------------

Let us now explore the implications of the NNLO self-spin correction to $r_c$. [We recall
that the corrections to  $r_c$ only enter the Hamiltonian, and do not concern the waveform.]
This
is done in Fig.~\ref{fig:Qomg_nnlo}, which refers to configurations {\tt SLy-q1.2-sA005-sB008},
i.e., an unequal-mass binary with physically motivated values of the spins and medium
values of $(C_{QA},C_{QB})$. The NNLO term yields an additional acceleration of the phasing.
However, the corresponding modification of the NLO curve is {\it smaller} than the modification
of the LO curve brought by the NLO contributions. [This intuitively suggests some type of
convergence of the EOB $Q_\omega$ curves as the amount of analytical information is increased.]
 Since the calculation of the self-spin terms in the energy flux (and waveform) is currently not
available at NNLO, we cannot include in Fig.~\ref{fig:Qomg_nnlo}  the corresponding
TaylorF2 curve. This raises the issue of knowing to which extent the NNLO curve represents a faithful representation
of the complete self-spin effects. We can venture an answer based on the knowledge of what
happens at NLO. Indeed, in the latter case one finds that the effect of the NLO waveform amplitude
terms is almost negligible: the $Q_\omega$ curve obtained by switching off these terms is 
essentially superposed to the one with the NLO waveform corrections. Based on this finding,
we expect that a similar situation will hold at the NNLO level.

Let us finally get an idea of the effect of the complete quartic-in-spin term once
included into $r_c$. This can be done evaluating numerically $C_\text{Oct}$
and $C_\text{Hex}$ using the quasi-universal fits of Eq.~(90) of Ref.~\cite{Yagi:2014bxa}.
Considering again the configuration {\tt SLy-q1.2-sA005-sB008}
of Fig.~\ref{fig:Qomg_nnlo} above, one notes that the NNLO spin-square effect is
determined by the coefficient $\delta a^2_{\rm NNLO}\simeq 0.108$ entering $r_c$.
For the same configuration, the quartic-in-spin LO coefficient of Eq.~\eqref{eq:delta_a4_LO},
that can be seen as a correction to $\delta a^2_{\rm NNLO}$, numerically reads
$\delta a^4_{\rm LO}\simeq 3.05\times 10^{-4}$, so that its effect would be completely
negligible on the phasing analysis of Fig.~\ref{fig:Qomg_nnlo}. One also
easily checks that one would need to have $\chi_A=\chi_B\approx 0.29$ so
to have $\delta a^4_{\rm LO}\approx 0.1$, thus yielding a phasing correction,
at the $Q_\omega$ level, comparable to the $\delta a^2_{\rm NNLO}$ displayed
in Fig.~\ref{fig:Qomg_nnlo}.

\section{Conclusions}
\label{sec:conclusions}
%======================
We have incorporated the EOS-dependent self-spin terms (or monopole-quadrupole effects)
in \TEOBResumS{} at NLO (i.e., at 3PN order, in both the Hamiltonian and the flux) and
at NNLO (4PN order, though only in the Hamiltonian, since the corresponding information
in the flux is not available yet). Following previous work~\cite{Nagar:2018zoe}, this was
done through a modification of the centrifugal radius function $r_c(r, \tilde a_i)$, which now 
depends on the spin-induced quadrupole moments. 

Our main findings can be summarized as follows:
\begin{itemize}
\item[(i)] Using the $Q_\omega$ gauge-invariant description of the phasing, we have found
that, once incorporated in the EOB formalism, NLO self-spin effects during the late inspiral 
are {\it more phase accelerating} than the LO ones (consistent with the corresponding PN behavior)
but at the same time different and {\it less phase accelerating} than the corresponding PN description
at NLO expressed through the TaylorF2 approximant. We have verified this to be the case
for a few (though illustrative) EOS choices and binary parameters
\item[(ii)] The resummed EOB self-spin phasing during the inspiral can be well approximated 
by augmenting the TaylorF2 approximant by the LO self-spin tail term. 
The resulting approximant delivers a simple phasing expression
that is consistent with the EOB one up to dimensionless frequency up to $M\omega\simeq 0.06$
\item[(iii)] In general, the fact that the PN prediction is always more phase accelerating than the EOB one
may have consequences on the estimate of these effects on real data, especially in the case
of fast spinning BNS~\cite{Harry:2018hke} made by recycled NS. This also indicates that current
waveform models, notably {\tt PhenomPv2\_NRTidal}~\cite{Dietrich:2018uni}, that incorporate
self-spin effects and  that have been used for the analysis of GW170817, should be updated
accordingly. This may eventually affect the evaluation of the systematics due to waveform
models in the analysis of  GW170817. 
\item[(iv)] Similarly, it will also be interesting to repeat the parameter estimation of
  GW170817 performed in Ref.~\cite{Abbott:2018wiz} using \TEOBResumS{} using the current version
  of the model that incorporates up to NNLO, EOS-dependent, self-spin effects. Note that this
  will also imply incorporating more, point-mass, spin-dependent terms than those currently
  present in the model.
\item[(v)] We have illustrated how to consistently compare the EOB and TaylorF2 phasing,
  notably in the low-frequency regime, using the $Q_\omega(\omega)$ function. In the present
   paper this comparison was restricted to the quadrupole-monopole part of the phasing.
  It would be interesting to generalize it to the other parts, so to have in hands precise comparisons between the orbital,
  spin-orbit or spin-spin parts. We postpone such a comparison to future work.
\end{itemize}

\begin{acknowledgments}
  F.~M. thanks IHES for hospitality during the final stages of development of this work.
\end{acknowledgments}

\appendix

\section{Cubic-in-spin terms within the EOB Hamiltonian}
\label{app:S3}
In this Appendix, we discuss preliminary ways of incorporating in the EOB Hamiltonian
the LO cubic-in-spin contributions to the dynamics derived in Ref.~\cite{Levi:2015uxa}.

As briefly mentioned in the main text, let us first recall that in \TEOBResumS{}
some contributions cubic in spin are already incorporated in the EOB Hamiltonian
via the presence of $u_c(r, \tilde a_i)$-dependent factors in the gyro-gravitomagnetic
functions $G_S$ and $G_{S_*}$  parametrizing the spin-orbit sector of the
Hamiltonian (see Sec.~\ref{sec:S2}). Indeed, we have $u_c^2=u^2( 1 + ({\rm spin-quadratic}) O(u^2))$
so that the linear-in-spin couplings defined by $G_S(u_c)$ and $G_{S_*}(u_c)$ automatically contain some $O(u^5)$
spin-cubic contributions. However, one checks (see main text, Eqs.~\eqref{EjS3}-\eqref{EjS3_LS})
that the spin-cubic terms thereby already incorporated in the Hamiltonian are not
the ones needed from the results of Ref.~\cite{Levi:2015uxa}.
This remark suggests a way of incorporating the needed spin-cubic terms in a resummed manner, namely to introduce new
definitions of the function $u_c(r, \tilde a_i)$ to be used as inputs in modified definitions of the 
gyro-gravitomagnetic functions $G_S$ and $G_{S_*}$. Say
\begin{align}
G_S     =&~ 2 u u_{c, G_S}^2 \hat{G}_S(u_c), \nonumber \\
G_{S_*} =&~ \frac{3}{2} (u_{c, G_{S_*}}^2)^{3/2} \hat{G}_{S_*}(u_c), \nonumber \\
\end{align}
where $\hat{G}_S(u_c)= 1+ O(u_c)$ and $\hat{G}_{S_*}=1+ O(u_c)$ are PN correcting factors \cite{Nagar:2011fx,Damour:2014yha}. [The arguments $u_c$ entering $\hat{G}_S(u_c)$ and  $\hat{G}_{S_*}(u_c)$ can be taken as being any
variable such that $u_c^2=u^2( 1 + ({\rm spin-quadratic}) O(u^2))$.]
We found that this possibility a priori involves six parameters, parametrizing the two different spin-quadratic expressions
separately entering the modified definitions of $r_{c, G_S}^2=r^2 + {\rm spin-quadratic}(1+ O(u)) $
and $r_{c, G_{S_*}}^2=r^2 + {\rm spin-quadratic}(1+ O(u)) $, modelled on Eqs. \eqref{rc2}, \eqref{aQ}. This leaves the
freedom to  arbitrarily choose two among these six parameters. This freedom of choice can be used to simplify
the resulting definitions. We have explored this avenue. However, at this stage we did not find a unique,
convincing way of simplifying the two spin-quadratic expressions entering $r_{c, G_S}^2$ and
$r_{c, G_{S_*}}^2$. We leave further studies along this avenue to future work.

Without committing ourselves to any specific resummed way of incorporating spin-cubic
terms in the EOB Hamiltonian, we wish, however, to display here the full information
needed for such definitions. To do this we will parametrize the spin-cubic contributions
in the following 4-parameter, {\it non-committal} form
\begin{align}
\label{Hso4param}
H_{\rm SO} =&~ p_{\varphi} \biggl[\tilde{G}(u,\tilde{a}_i ) + \bigl(b_{30} \tilde{a}_{A}^3 + b_{21} \tilde{a}_{A}^2 \tilde{a}_{B}  \nonumber 	\\
&+b_{12} \tilde{a}_{A} \tilde{a}_{B}^2 +b_{03} \tilde{a}_{B}^3\bigr)u^5\biggr],
\end{align}
where
\begin{equation}
\tilde{G}(u,\tilde{a}_i ) = 2 u^3 \hat{G}_S(u) \hat S + \frac{3}{2} u^3  \hat{G}_{S_*}(u) \hat S_*.
\end{equation}

We then follow the procedure described in Section~\ref{sec:summary} to determine the four
parameters $ b_{30} , b_{21} , b_{12} , b_{03} $ entering this parametrization.
By calculating the corresponding modified version of  $c_{7}^{S^3_{\rm LO}}$ and comparing it to Eq.~\eqref{EjS3}, we obtain the simple expressions
\begin{align}
b_{30} =&~ C_{\text{Oct} A} - 3C_{Q A}, \nonumber \\
b_{21} =&~ -6, \nonumber \\
b_{12} =&~ -6, \nonumber \\
b_{03} =&~ C_{\text{Oct} B} - 3C_{Q B}.
\end{align}

Let us note that if we consider the BH limit where $C_{Q}=C_\text{Oct}=1 $ the needed modified
spin-orbit coupling takes the very simple form
\begin{equation}
H_{\rm SO \,  BBH} = p_{\varphi} \left[ \tilde{G}(u) - 2 (\hat{S} + \hat{S}_*)^3 u^5  \right].
\end{equation}

Let us also note that if we insist on utilizing the full \TEOBResumS{} structure, keeping the cubic-in-spin terms that come from the use of $r_c(r, \tilde a_i)$, as defined in Eq. \eqref{rc2} above, we must modify the expressions
of the parameters $ b_{30} , b_{21} , b_{12} , b_{03} $ into
\begin{align}
b'_{30} =&~ C_{\text{Oct} A} - \frac{7}{8}C_{Q A}- \frac{1}{8}X_{AB}C_{Q A}, \nonumber \\
b'_{21} =&~ -\frac{7}{4} +\frac{17}{8}C_{Q A} - X_{AB} \left(\frac{1}{4} - \frac{1}{8}C_{Q A}\right), \nonumber \\
b'_{12} =&~ -\frac{7}{4} +\frac{17}{8}C_{Q B} + X_{AB} \left(\frac{1}{4} - \frac{1}{8}C_{Q B}\right), \nonumber \\
b'_{03} =&~ C_{\text{Oct} B} - \frac{7}{8}C_{Q B} + \frac{1}{8}X_{AB}C_{Q B}.
\end{align} 
In that case, the BBH case leads to the following very simple correction to the spin-sector implied
by the current \TEOBResumS{} model:
\begin{equation}
H_{\rm SO \,  BBH} = p_{\varphi} \left[ \tilde{G} + \frac{1}{4} (\hat{S} + \hat{S}_*)^2 \hat{S}_* u^5\right].
\end{equation}

We postpone a comparison of the various avenues mentioned here to a later work.

%-------------------------------------------------------------
\section{Post-adiabatic dynamics}
\label{sec:PA}

The EOB/PN comparisons done in the main text employ
\TEOBResumS{}~{\tt v0.1}, that was implemented in C++.
An equivalent, though tidally enhanced model
(see~\cite{Akcay:2018yyh})
and computationally more efficient version of the model
(implemented in C) is {\tt v1.0}. All our codes are publicly
available at
\begin{center}
	{\footnotesize \url{https://bitbucket.org/account/user/eob_ihes/projects/EOB}}
\end{center}

\TEOBResumS{} {\tt v1.0}  optionally implements the post-adiabatic (PA)
approximation to efficiently deal with the long inspiral phase~\cite{Nagar:2018gnk}.
Following the logic of Appendix B of Ref.~\cite{Akcay:2018yyh},
we here present the performance of the PA evolution in the case of
spinning neutron stars. The result presented here are obtained
incorporating (i) NLO spin-quadratic information in the waveform
and (ii) NNLO spin-quadratic information in the Hamiltonian.
This should be considered as the default choice in \TEOBResumS{}
for what concerns spinning BNS. Optionally, it is possible to
switch on the EOS-dependence in the quartic-in-spin correction
to $r_c$, Eq.~\eqref{eq:delta_a4_LO}, though this does not come
as default choice in the code.

Within \TEOBResumS{}, the dynamics of a binary system
is usually determined by numerically solving the four Ordinary Differential
Equations (ODEs) of the Hamiltonian relative dynamics.
The time needed to solve these ODEs usually weighs
as the main contribution to the waveform evaluation time.
Using this C-implementation of \TEOBResumS{}, a typical
time-domain BNS waveform requires $\sim 1$~sec to be generated
starting from a GW frequency of 10~Hz by means of standard
Runge-Kutta integration routines with adaptive step-size that
are publicly available through the GNU Scientific Library (\texttt{GSL}).
Thus, ODE integration as is cannot be used in parameter estimation
runs that require the generation of $~ 10^7$ waveforms. 
Reference~\cite{Nagar:2018gnk} proposed a method of reducing the evaluation
time by making use of the PA approximation to compute the system
dynamics. In Ref.~\cite{Akcay:2018yyh} we restricted to non spinning
BNS and we showed, for the first time, how a waveform obtained from
the complete ODE evolution compares with a waveform obtained by
stitching the PA dynamics to the complete dynamics for the last few
orbits up to merger (as suggested in~\cite{Nagar:2018gnk}). 
For completeness, we here present the same comparison for 
two illustrative, spinning, BNS systems.

\begin{figure}[t]
\vspace{2mm}
\centering
\includegraphics[width=0.45\textwidth]{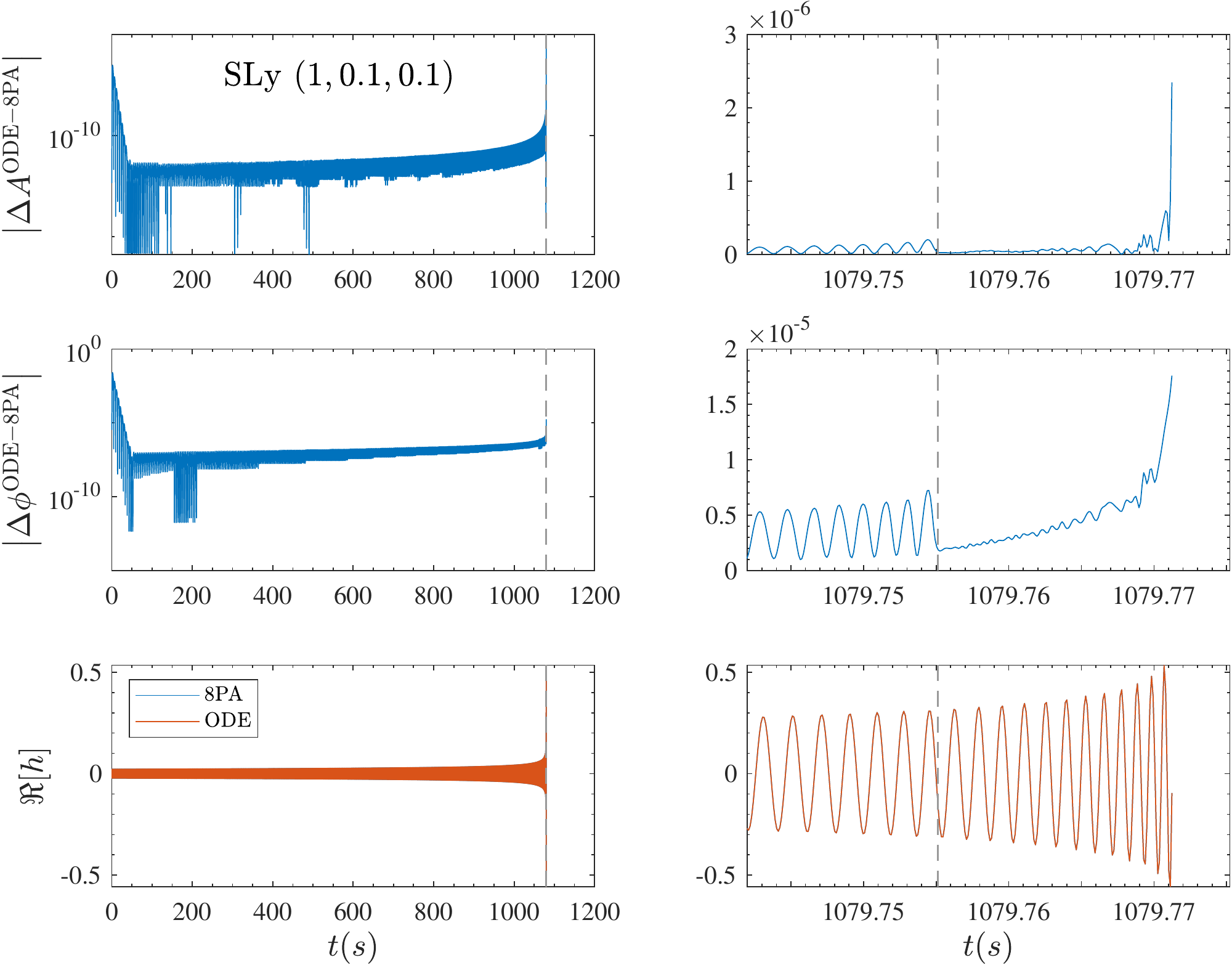}\\
\includegraphics[width=0.45\textwidth]{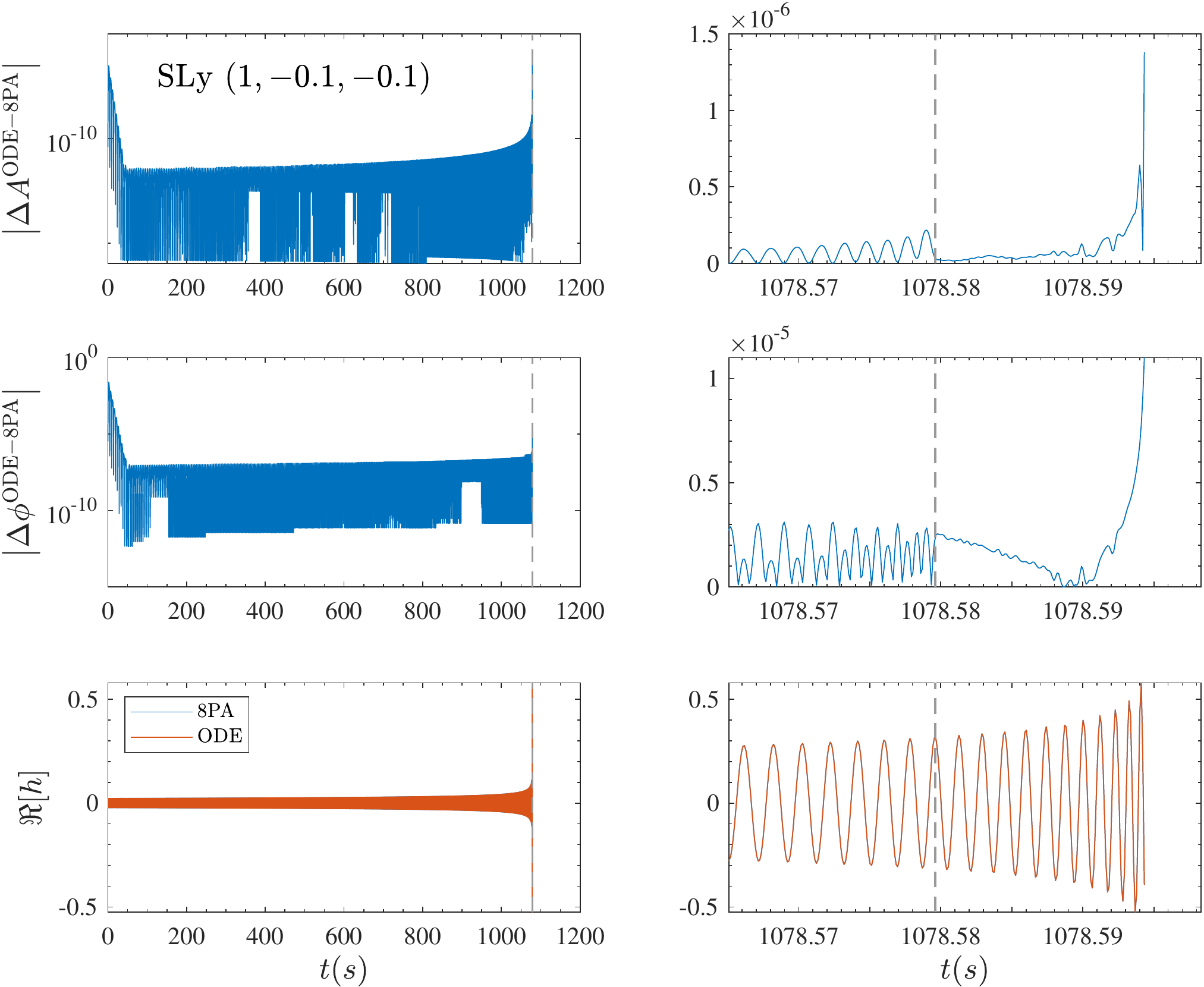}
\caption{
	\label{fig:PA}
	Comparison between the waveforms (obtained summing all modes up to $\ell_{\rm max}=8$,
        see Eq.~\eqref{eq:hpc}) computed solving the ODEs with the \texttt{GSL} \texttt{rk8}
        routine with adaptive stepsize and the PA waveform completed with
	the same ODE solver after $r<r_\text{min}$.
	We have considered BNS systems with $1.35M_\odot+1.35M_\odot$
	and SLy EOS (see first row of Table~\ref{tab:RunsTable}),
        starting at initial frequency 10~Hz.
	Dimensionless spins are $\chi_A = \chi_B = 0.1$ (top) and $\chi_A = \chi_B = -0.1$ (bottom).  
	The parameters used for the PA run are listed in Table~\ref{table:PAtime}.
	The dashed grey line marks the stitching point, $r_\text{min}$,
        between the PA and ODE-based dynamics.
        Given the waveform strain as $h\equiv A
	e^{\rm -i \phi}$, we defined the phase difference as
	$\Delta\phi^{\rm ODE-8PA} \equiv \phi^{\rm ODE} -
	\phi^{\rm 8PA}$ and the fractional amplitude
	difference as $\Delta A^{\rm ODE-8PA} \equiv (A^{\rm
		ODE} - A^{\rm 8PA})/A^{\rm ODE}$. 
	The larger differences at the beginning of the evolution
        are partly due to the fact that the complete ODE 
        is started using only 2PA-accurate initial data.}
\end{figure}

Let us briefly summarize the approach of Ref.~\cite{Nagar:2018gnk}.
The PA approximation to the EOB dynamics was introduced in
Refs.~\cite{Buonanno:1998gg,Buonanno:2000ef} (and expanded in
Refs.~\cite{Damour:2007yf,Damour:2012ky}) and is currently used
to initialize the relative dynamics in \TEOBResumS{} with negligible
eccentricity. Using this approximation, it is possible to analytically
compute the radial and angular momentum of a binary system,
under the assumption that the radiation reaction force is small.
This is true in the early inspiral phase and progressively
loses validity when the two objects get close. The approach starts by
considering the conservative system, when the flux is null, and then
computes the successive corrections to the momenta. We denote as
$n$PA the $n$-th order iteration of this procedure.
%===========================================
% Table of performance without interpolation
%===========================================
\begin{table}[h]
\centering
\begin{tabular}{ccccccc}
	\hline\hline
	$f_0$ [Hz] & $r_0$ & $r_{\rm min}$ & $N_r$ & $\Delta r$ & $\tau_{\rm 8PA}$ [sec] & $\tau_{\rm ODE}$ [sec] \\
	\hline
	20 & 112.81 & 12 & 500 & 0.20 & 0.03 & 0.53 \\
	10 & 179.02 & 12 & 830 & 0.20 & 0.05 & 1.1  \\
	\hline\hline
\end{tabular}
\caption{
	\label{table:PAtime}
	Performance of \TEOBResumS{}~{\tt v1.0}
        for a BNS system with $1.35M_\odot+1.35M_\odot$, SLy EOS and
        $\chi_A = \chi_B = 0.1$. The waveform for the $\chi_A = \chi_B = -0.1$
        case is a little shorter but the evaluation times are comparable to
        the ones showed in the table. $f_0$ and $r_0$ denote the initial
        GW frequency and radial separation. The 8PA dynamics is computed
        on a grid with $N_r$ points and grid separation $\Delta r$
	that ends at $r_{\rm min}$ and then is completed by the standard ODE one. The evaluation
	times $\tau$ are determined using a standard Intel Core i7, 1.8GHz and 16GB RAM.
	The code is compiled with the GNU gcc compiler using O3 optimization.}
\end{table}
Practically, to compute the PA dynamics, we first build a uniform
radial grid from the initial radius $r_0$ to an $r_\text{min}$ up until
which we are sure the approximation holds. 
We then analytically compute the momenta that correspond to each radius
at a chosen PA order. Finally, we determine the full dynamics recovering
the time and orbital phase by quadratures.
From $r_\text{min}$ we can then start the usual ODE-based dynamics
using the PA quantities as initial data as it is usually done
(at 2PA order) in \TEOBResumS{}. With this method one can avoid
to numerically solving two Hamilton equations (those for the momenta),
while the orbital phase and time can be obtained by quadratures over
a rather sparse radial grid.
%===========================================
% Table of performance with interpolation
%===========================================
%\begin{table}[h!]
\begin{table}[t]
\centering
\begin{tabular}{cccc}
	\hline\hline
	$f_0$ [Hz] & $r_0$ & $\tau_{\rm 8PA}^{\rm int}$ [sec] & $\tau_{\rm ODE}^{\rm int}$ [sec] \\
	\hline
	20 & 112.81 & 0.10 & 0.64 \\
	10 & 179.02 & 0.37 & 1.70 \\
	\hline\hline
\end{tabular}
\caption{
	\label{table:PAtimeint}
	Performance of the \TEOBResumS{}~{\tt v1.0} when the final waveform is interpolated
    on a time grid evenly sampled at 1/(4096~Hz). We use the standard spline interpolation
    routine implemented in the \texttt{GSL} library. The considered system coincides with the one of Table~\ref{table:PAtime}.}
\end{table}

Figure~\ref{fig:PA} displays the performance of the PA approximation (at 8PA order) for two,
illustrative, spinning BNS systems  with $1.35M_\odot+1.35M_\odot$ and SLy EOS.
The figure shows the distance-normalized waveform strain $h\equiv {\cal R}(h_+-\ii h_\times)$,
where we recall that the multipolar decomposition of the waveform reads
\be
\label{eq:hpc}
{\cal R}(h_+-\ii h_\times)=\sum_{\ell =2}^{\ell_{\max}}\sum_{m=-\ell}^{\ell}h_{\ell m}\,{}_{-2}Y(\theta,\Phi),
\ee
where $h_{\ell m}$ are the waveform (complex) multipoles and ${}_{-2}Y(\theta,\Phi)$ are
the $s=-2$ spin-weighted spherical harmonics (that are here evaluated at $\theta=\Phi=0$).
In fig.~\ref{fig:PA} we evaluate $h$ with $\ell_{\max}=8$, i.e. retaining in the waveform
the same 35 multipoles that are used to compute the EOB radiation reaction.
For each binary, each subpanel displays the the waveform fractional-amplitude
difference (top), phase difference (medium) and real part of the waveform strain.
The left columns offer a global view, while the right columns focus on the last
few GW cycles up to merger. The vertical dashed line marks the time where
the PA evolution is stitched to the ODE evolution for the last orbits where the
PA approximation breaks down. Table~\ref{table:PAtime} highlights the performances
of \TEOBResumS{}~{\tt v1.0} for such a case.
Note that the initial radius is determined by solving the circular Hamilton's
equations instead of relying on the Newtonian Kepler's law, as discussed
in Sec. VI of Ref.~\cite{Nagar:2018zoe}.

The waveform computed using the PA dynamics (completed with the ODE for the last few orbits)
only takes around 50 milliseconds to be evaluated. Such a time is comparable to 
the one needed by the surrogate models that are currently being constructed in order
to reduce waveform evaluation times (see e.g.~\cite{Lackey:2016krb})
and that, typically, only involve the $\ell=m=2$ mode.
Finally, Table~\ref{table:PAtimeint} illustrates the performance
of \TEOBResumS{} when the waveform of above, which is obtained
on a {\it nonuniform} temporal grid (because the corresponding radial
grid is evenly spaced), is interpolated on an evenly spaced time grid,
sampled at $\Delta t^{-1}=4096$~Hz, that is usually neecessary to compute
the Fourier transform with standard algorithms.
It is remarkable that the generation time of the {\it full multipolar waveform}
is below 1~sec {\it also} when the starting frequency is 10~Hz.
Such interpolation is done with the spline interpolant that is freely available in
the \texttt{GSL} library and looks to be the main routine responsible for the computational
cost of the waveform generation. We expect this can be further speed
up exploting vectorization or shared memory parallelization.
Similarly, one expects that the number of radial gridpoints needed might be lowered
further by adopting a quadrature formula at higher order (now a third-order one is
implemented, following Ref.~\cite{Nagar:2018gnk}) to recover the orbital phase and time.
Such technical improvements will be explored extensively in forthcoming works.

%---------------------------------------
\section{Using $\hat{S}$ and $\hat{S}_*$ as spin variables.}
\label{app:S_conv}
For completeness, we report here the results of Sec.~\ref{sec:summary}
using $\hat{S}$ and $\hat{S}_*$ as spin variables. First, the newly
computed $\delta a_{\rm NLO}^2$ and $\delta a_{\rm NNLO}^2$ are written
as quadratic forms in $(S,S_*)$ as 
\begin{widetext}
\begin{align}
\delta a^2_{\rm NLO} =& \frac{1}{1-4\nu}\Bigg\{\left[-4+9\nu+\left(2-5\nu\right)\left(C_{QA}+C_{QB}\right)+\left(2-\nu\right)X_{AB}\left(C_{QA}-C_{QB}\right) \right]\hat{S}^2 \nonumber \\
&+\left[-\frac{9}{2}+11\nu+\left(1-\nu\right)\left(C_{QA}+C_{QB}\right)-\left(1+\nu\right)X_{AB}\left(C_{QA}-C_{QB}\right) \right]\hat{S}_*^2\nonumber \\
&+\left[-2+22\nu-6\nu\left(C_{QA}+C_{QB}\right)-2\nu X_{AB}\left(C_{QA}-C_{QB}\right)\right]\hat{S}\hat{S}_*\Bigg\},\\
\delta a_{\rm NNLO}^2 =&  \frac{1}{1-4\nu} \Bigg\{\Big[-\frac{275}{14}+\frac{561}{14}\nu+\frac{675}{14}\nu^2\nonumber \\
&+\left(\frac{275}{28}-\frac{1633}{56}\nu+\frac{207}{28}\nu^2\right)\left(C_{QA}+C_{QB}\right)
+\left(\frac{275}{28}-\frac{533}{56}\nu\right)X_{AB}\left(C_{QA}-C_{QB}\right) \Big]\hat{S}^2\nonumber \\
& +\left[-\frac{153}{8}+\frac{173}{4}\nu+\frac{381}{14}\nu^2+\left(4-\frac{47}{8}\nu+\frac{207}{28}\nu^2\right)\left(C_{QA}+C_{QB}\right)-\left(4+\frac{17}{8}\nu\right)X_{AB}\left(C_{QA}-C_{QB}\right) \right]\hat{S}_*^2\nonumber \\
&+\left[-\frac{25}{2}+\frac{4727}{56}\nu-\frac{163}{14}\nu^2-\left(\frac{387}{14}-\frac{207}{14}\nu\right)\nu\left(C_{QA}+C_{QB}\right)-\frac{163}{14}\nu X_{AB}\left(C_{QA}-C_{QB}\right)\right]\hat{S}\hat{S}_*\Bigg\}.
\end{align}
\end{widetext}
Note that the use of $(S,S_*)$ leads to formally singular terms when $\nu=1/4$. 
This singularity is actually reabsorbed by $(\hat{S},\hat{S}_*)$
when the limit is done carefully taking into account the various mass terms.
From these equations, one can obtain the orbital angular momentum $j$ as a function of $u$, namely
\begin{widetext}
\begin{align}
j(u)=&~
\frac{1}{\sqrt{u}}+\frac{3}{2}\sqrt{u}
-3\left(\hat{S} + \frac{3}{4}\hat{S}_*\right)u\nonumber\\
&+\Bigg\{-\frac{27}{8}-\frac{3}{2}\nu+\frac{1}{1-4\nu}\left[-2\nu+\left(\frac{1}{2}-\nu\right)\left(C_{QA}+C_{QB}\right)+\frac{1}{2}X_{AB}\left(C_{QA}-C_{QB}\right) \right]\hat{S}^2 \nonumber \\
&+\frac{1}{1-4\nu}\left[-2\nu+\left(\frac{1}{2}-\nu\right)\left(C_{QA}+C_{QB}\right)-\frac{1}{2}X_{AB}\left(C_{QA}-C_{QB}\right) \right]\hat{S}_*^2 + \nonumber \\
&+\frac{2}{1-4\nu}\left[1-2\nu-\nu(C_{QA}+C_{QB})\right]\hat{S}\hat{S}_*\Bigg\}u^{3/2}
+\left[\left(-\frac{15}{2}+\frac{5}{4}\nu\right)\hat{S}+\left(-\frac{27}{8}+\frac{3}{2}\nu\right)\hat{S}_*\right]u^2\nonumber\\
&+\biggl\{\frac{135}{16}+\left(-\frac{433}{12}+\frac{41}{32}\pi^2\right)\nu\nonumber \\
&+\frac{1}{1-4\nu}\biggl[-\frac{1}{2}-\frac{31}{4}\nu+\left(\frac{11}{4}-\frac{27}{4}\nu\right)\left(C_{QA}+C_{QB}\right) +\left(\frac{11}{4}-\frac{5}{4}\nu\right)X_{AB}\left(C_{QA}-C_{QB}\right) \biggr]\hat{S}^2 \nonumber \\
&+\frac{1}{1-4\nu}\left[-\frac{99}{32}+\frac{21}{8}\nu+\left(\frac{3}{2}-\frac{7}{4}\nu\right)\left(C_{QA}+C_{QB}\right)-\left(\frac{3}{2}+\frac{5}{4}\nu\right)X_{AB}\left(C_{QA}-C_{QB}\right) \right]\hat{S}_*^2\nonumber \\
&+\frac{1}{1-4\nu}\left[\frac{21}{4}-\frac{3}{2}\nu-\frac{17}{2}\nu\left(C_{QA}+C_{QB}\right)-\frac{5}{2}\nu\left(C_{QA}-C_{QB}\right)\right]\hat{S}\hat{S}_*\biggr\}u^{5/2} \nonumber\\
&+\left[\left(-\frac{189}{8}+35\nu+\frac{5}{16}\nu^2\right)\hat{S}+\left(-\frac{63}{8}+\frac{225}{8}\nu+\frac{15}{32}\nu^2\right)\hat{S}_*\right]u^3\nonumber\\
&+\Biggl\{\frac{2835}{128}-\left(\frac{3029}{120} + 32 \gamma + \frac{3503}{2048}\pi^2 + 64 {\rm log}(2) + 16 {\rm log}(u)  \right)\nu + \left(\frac{539}{12}-\frac{205}{128}\pi^2\right)\nu^2 \nonumber \\
&+\frac{1}{1-4\nu}\biggl[\frac{67}{14}-\frac{3459}{56}\nu+\frac{2319}{28}\nu^2+\left(\frac{1055}{112}-\frac{3317}{112}\nu+\frac{495}{56}\nu^2\right)\left(C_{QA}+C_{QB}\right)\nonumber \\
&+\left(\frac{1055}{112}-\frac{1207}{112}\nu\right)X_{AB}\left(C_{QA}-C_{QB}\right) \biggr]\hat{S}^2 +\frac{1}{1-4\nu}\biggl[-\frac{585}{64}+\frac{35}{16}\nu+\frac{1395}{28}\nu^2\nonumber \\
&+\left(\frac{49}{16}-\frac{67}{16}\nu+\frac{495}{56}\nu^2\right)\left(C_{QA}+C_{QB}\right)-\left(\frac{49}{16}+\frac{31}{16}\nu\right)X_{AB}\left(C_{QA}-C_{QB}\right) \biggr]\hat{S}_*^2\nonumber \\
&+\frac{1}{1-4\nu}\biggl[\frac{113}{8}-\frac{2545}{56}\nu+\frac{918}{7}\nu^2 - \left(\frac{699}{28}-\frac{495}{28}\nu\right)\nu\left(C_{QA}+C_{QB}\right) \nonumber \\
&-\frac{89}{7}\nu X_{AB}\left(C_{QA}-C_{QB}\right) \biggr]\hat{S}\hat{S}_*\Biggr\}u^{7/2} + \O\left[u^4\right].
\end{align}
\end{widetext}
Similarly, the inverse expression can be written as
\begin{widetext}
\begin{align}
u(j)=&~\frac{1}{j^2}
+\frac{3}{j^4}
-6\left(\hat{S} + \frac{3}{4}\hat{S}_*\right)\frac{1}{j^5} \nonumber \\
&+\Bigg\{18-3\nu+\frac{1}{1-4\nu}\left[-4\nu+\left(1-2\nu\right)\left(C_{QA}+C_{QB}\right)+X_{AB}\left(C_{QA}-C_{QB}\right) \right]\hat{S}^2 \nonumber \\
&+\frac{1}{1-4\nu}\left[-4\nu+\left(1-2\nu\right)\left(C_{QA}+C_{QB}\right)-X_{AB}\left(C_{QA}-C_{QB}\right) \right]\hat{S}_*^2 + \nonumber \\
&+\frac{4}{1-4\nu}\left[1-2\nu-\nu(C_{QA}+C_{QB})\right]\hat{S}\hat{S}_*\Bigg\}\frac{1}{j^6}
+\left[\left(-69+\frac{5}{2}\nu\right)\hat{S}+\left(-\frac{189}{4}+3\nu\right)\hat{S}_*\right]\frac{1}{j^7}\nonumber\\
&+\biggl\{135+\left(-\frac{311}{3}+\frac{41}{16}\pi^2\right)\nu \nonumber \\
&+\frac{1}{1-4\nu}\left[62-\frac{619}{2}\nu+\left(16-\frac{69}{2}\nu\right)\left(C_{QA}+C_{QB}\right)+\left(16-\frac{5}{2}\nu\right)X_{AB}\left(C_{QA}-C_{QB}\right) \right]\hat{S}^2 \nonumber \\
&+\frac{1}{1-4\nu}\left[\frac{117}{4}-\frac{375}{2}\nu+\left(\frac{27}{2}-\frac{49}{2}\nu\right)\left(C_{QA}+C_{QB}\right)-\left(\frac{27}{2}+\frac{5}{2}\nu\right)X_{AB}\left(C_{QA}-C_{QB}\right) \right]\hat{S}_*^2\nonumber \\
&+\frac{1}{1-4\nu}\left[147-465\nu+-59\nu\left(C_{QA}+C_{QB}\right)-5\nu X_{AB}\left(C_{QA}-C_{QB}\right)\right]\hat{S}\hat{S}_*\biggr\}\frac{1}{j^8} \nonumber\\
&+\left[\left(-\frac{3069}{4}+172\nu+\frac{5}{8}\nu^2\right)\hat{S}+\left(-\frac{2007}{4}+\frac{585}{4}\nu+\frac{15}{16}\nu^2\right)\hat{S}_*\right]\frac{1}{j^9} \nonumber \\
&+\Biggl\{1134-\left(\frac{163063}{120}+64 \gamma- \frac{31921}{1024}\pi^2 + 128 {\rm log}(2) + 64 {\rm log}\left(1/j\right) \right)\nu + \left(\frac{1321}{12}- \frac{205}{64}\pi^2 \right)\nu^2 \nonumber \\
&+\frac{1}{1-4\nu}\biggl[\frac{19223}{14}-\frac{88953}{14}\nu+\frac{6855}{14}\nu^2+\left(\frac{5725}{28}-\frac{26753}{56}\nu+\frac{1251}{28}\nu^2\right)\left(C_{QA}+C_{QB}\right)\nonumber \\
&+\left(\frac{5725}{28}-\frac{3853}{56}\nu\right)X_{AB}\left(C_{QA}-C_{QB}\right) \biggr]\hat{S}^2 +\frac{1}{1-4\nu}\biggl[\frac{4653}{8}-\frac{12659}{4}\nu+\frac{5553}{14}\nu^2\nonumber \\
&+\left(158-\frac{2335}{8}\nu+\frac{1251}{28}\nu^2\right)\left(C_{QA}+C_{QB}\right)-\left(158+\frac{193}{8}\nu\right)X_{AB}\left(C_{QA}-C_{QB}\right) \biggr]\hat{S}_*^2\nonumber \\
&+\frac{1}{1-4\nu}\biggl[\frac{5119}{2}-\frac{503105}{56}\nu+\frac{12555}{14}\nu^2-\left(\frac{10149}{14}-\frac{1251}{14}\nu\right)\nu\left(C_{QA}+C_{QB}\right)\nonumber \\
&-\frac{1301}{14}\nu X_{AB}\left(C_{QA}-C_{QB}\right) \biggr]\hat{S}\hat{S}_* \Biggr\}\frac{1}{j^{10}} +\O\left[j^{-11}\right].
\end{align}
\end{widetext}

Finally, the gauge-invariant link between the binding energy and
the orbital angular momentum becomes
\begin{widetext}
\begin{align}
E_b(j)=&-\frac{1}{2 j^2} \Bigg\{1
+\frac{1}{4}(9+\nu)\frac{1}{j^2}
-4\left(\hat{S} + \frac{3}{4}\hat{S}_*\right) \frac{1}{j^3}\nonumber\\
&+\Bigg\{\frac{1}{8}\left(81-7\nu+\nu^2\right)+\frac{1}{1-4\nu}\left[-2\nu+\left(\frac{1}{2}-\nu\right)\left(C_{QA}+C_{QB}\right)+\frac{1}{2}X_{AB}\left(C_{QA}-C_{QB}\right) \right]\hat{S}^2 \nonumber \\
&+\frac{1}{1-4\nu}\left[-2\nu+\left(\frac{1}{2}-\nu\right)\left(C_{QA}+C_{QB}\right)-\frac{1}{2}X_{AB}\left(C_{QA}-C_{QB}\right) \right]\hat{S}_*^2 + \nonumber \\
&+\frac{2}{1-4\nu}\left[1-2\nu-\nu(C_{QA}+C_{QB})\right]\hat{S}\hat{S}_*\Bigg\}\frac{1}{j^4}
+\left[\left(-36-\frac{3}{4}\nu\right)\hat{S}+-\frac{99}{4}\hat{S}_*\right]\frac{1}{j^5}\nonumber\\
&+\biggl\{\frac{3861}{64}-\left(\frac{8833}{192}-\frac{41}{32}\pi^2\right)\nu-\frac{5}{32}\nu^2+\frac{5}{64}\nu^3 \nonumber \\
&+\frac{1}{1-4\nu}\left[32-152\nu-\nu^2+\left(\frac{25}{4}-\frac{53}{4}\nu-\frac{\nu^2}{2}\right)\left(C_{QA}+C_{QB}\right)+\left(\frac{25}{4}-\frac{3}{4}\nu\right)X_{AB}\left(C_{QA}-C_{QB}\right) \right]\hat{S}^2 \nonumber \\
&+\frac{1}{1-4\nu}\left[\frac{63}{4}-87\nu-\nu^2+\left(\frac{21}{4}-\frac{37}{4}\nu-\frac{\nu^2}{2}\right)\left(C_{QA}+C_{QB}\right)-\left(\frac{21}{4}+\frac{5}{4}\nu\right)X_{AB}\left(C_{QA}-C_{QB}\right) \right]\hat{S}_*^2\nonumber \\
&+\frac{1}{1-4\nu}\left[69-227\nu-2\nu^2-\left(23+\nu\right)\nu\left(C_{QA}+C_{QB}\right)-2\nu X_{AB}\left(C_{QA}-C_{QB}\right)\right]\hat{S}\hat{S}_*\biggr\}\frac{1}{j^6} \nonumber\\
&+\left[\left(-\frac{1701}{8}+\frac{3597}{4}\nu-195\nu^2\right)\hat{S}+\left(-\frac{4295}{16}+\frac{411}{8}\nu-\frac{5}{16}\nu^2\right)\hat{S}_*\right]\frac{1}{j^7}\nonumber \\
&+\Biggl\{\frac{53703}{128}-\left(\frac{989911}{1920}+\frac{128}{5} \gamma -\frac{6581}{512}\pi^2 +\frac{256}{5} {\rm log}(2) + \frac{128}{5} {\rm log}\left(1/j\right) \right)\nu + \left(\frac{8875}{384}- \frac{41}{64}\pi^2 \right)\nu^2 \nonumber \\
&-\frac{3}{64}\nu^3 +\frac{7}{128}\nu^4+ \frac{1}{1-4\nu}\Big[\frac{8013}{14}-\frac{71545}{28}\nu+\frac{2575}{28}\nu^2-\frac{3}{4}\nu^3+\left(\frac{7449}{112}-\frac{17119}{112}\nu +\frac{975}{112}\nu^2-\frac{3}{8}\nu^3\right)\left(C_{QA}+C_{QB}\right)\nonumber \\
&+\left(\frac{7449}{112}-\frac{2221}{112}\nu-\frac{5}{16}\nu^2\right)X_{AB}\left(C_{QA}-C_{QB}\right) \Big]\hat{S}^2 +\frac{1}{1-4\nu}\Big[252-\frac{10203}{8}\nu+\frac{2757}{28}\nu^2-\frac{3}{4}\nu^3\nonumber \\
&+\left(\frac{819}{16}-\frac{1473}{16}\nu+\frac{1199}{112}\nu^2-\frac{3}{8}\nu^3\right)\left(C_{QA}+C_{QB}\right)-\left(\frac{819}{16}+\frac{165}{16}\nu+\frac{11}{16}\nu^2\right)X_{AB}\left(C_{QA}-C_{QB}\right) \Big]\hat{S}_*^2\nonumber \\
&+\frac{1}{1-4\nu}\biggl[\frac{4041}{4}-\frac{201927}{56}\nu+\frac{5661}{28}\nu^2-\frac{3}{2}\nu^3 - \left(\frac{6591}{28}-\frac{533}{28}\nu+\frac{3}{4}\nu^2\right)\nu\left(C_{QA}+C_{QB}\right)\nonumber \\
&-\left(\frac{429}{14}+\nu\right)\nu X_{AB}\left(C_{QA}-C_{QB}\right)\biggr]\hat{S}\hat{S}_* \Biggr\}\frac{1}{j^{8}} + \O\left[j^{-9}\right]\Bigg\}.
\end{align}
\end{widetext}

\bibliography{refs20181219.bib,local.bib}

\end{document}